\documentclass[12pt]{article}
 \pdfoutput=1
\usepackage{latexsym}
\usepackage{amssymb}
\usepackage{amsmath}
\usepackage{dsfont} 
\usepackage{graphicx}
\usepackage{bbm}
\usepackage{color}
\input{colordvi} 

\newcommand{\comment}[1]{}
\def\tr{\operatorname{tr}}

\def\e{\mathrm{e}}
\def\im{\mathrm{i}}

\def\Re{\mathrm{Re}\,}
\def\Im{\mathrm{Im}\,}
\def\com{\mathrm{Com}}

\def\e{{\mathrm e}}
\newcommand{\lsp}{\mathrm{span}}
\newcommand{\ran}{\mathop{\mathrm{ran}}}
\newcommand{\id}{\mathbbm{1}}
\newtheorem{thm}{Theorem}
\newtheorem{lem}[thm]{Lemma}
\newtheorem{prop}[thm]{Proposition}
\newtheorem{cor}[thm]{Corollary}
\newtheorem{exa}{Example}
\newtheorem{rem}{Remark}
\newtheorem{hyp}{Hypothesis}
\newtheorem{hypprime}{Hypothesis}
\newtheorem{definition}{Definition}

\newcommand{\ket}[1]{\left| #1 \right\rangle}

\title{Adiabatic theorems for generators of contracting evolutions}
\author{J.E. Avron, M. Fraas,
\\
\small{Department of Physics, Technion, 32000 Haifa, Israel}
\\ G.M. Graf, P. Grech
\\
\small{Theoretische Physik, ETH Zurich, 8093 Zurich, Switzerland} }

\begin{document}

\maketitle
\begin{abstract}
We develop an adiabatic theory for generators of contracting evolution on Banach spaces. This provides a uniform framework for a host of adiabatic theorems ranging from unitary quantum evolutions through quantum evolutions of open systems generated by Lindbladians all the way to classically driven stochastic systems. In all these cases the adiabatic evolution approximates, to lowest order, the natural notion of parallel transport in the manifold of instantaneous stationary states. The dynamics in the manifold of instantaneous stationary states and transversal to it have distinct characteristics: The former is irreversible and the latter is transient in a sense that we explain. Both the gapped and gapless cases are considered. Some applications are discussed.
\end{abstract}

\section{Introduction}

We develop a framework for the adiabatic theory of systems whose evolution is governed by a slowly evolving family of linear operators generating a
contraction in a Banach space \cite{NR, AS, J}. More precisely, we study equations of the form

\begin{equation} \label{adeq}
 \varepsilon \dot{x}(s) = L(s) x(s),
\end{equation}
where $L(s)$ is, for any fixed $s$, the generator of a contraction semigroup.

The framework encompasses a wide range of applications from driven stochastic
systems generated in a Markovian process, through isolated quantum systems
undergoing unitary evolution generated by Hamiltonians, culminating in open
quantum systems whose evolution is generated by Lindblad operators. 

Adiabatic evolutions have a geometric character. As we shall see,
the manifold of instantaneous
stationary vectors, namely $\ker L(s)$, has a distinguished complement, with the property that 
a vector near the former evolves with a velocity in the latter, to 
leading order in $\varepsilon$.
Hence, to lowest 
order in the adiabatic limit, the vector is parallel transported with the 
manifold. 

\begin{figure}[htb]
\centering
\includegraphics[width=0.5\textwidth]{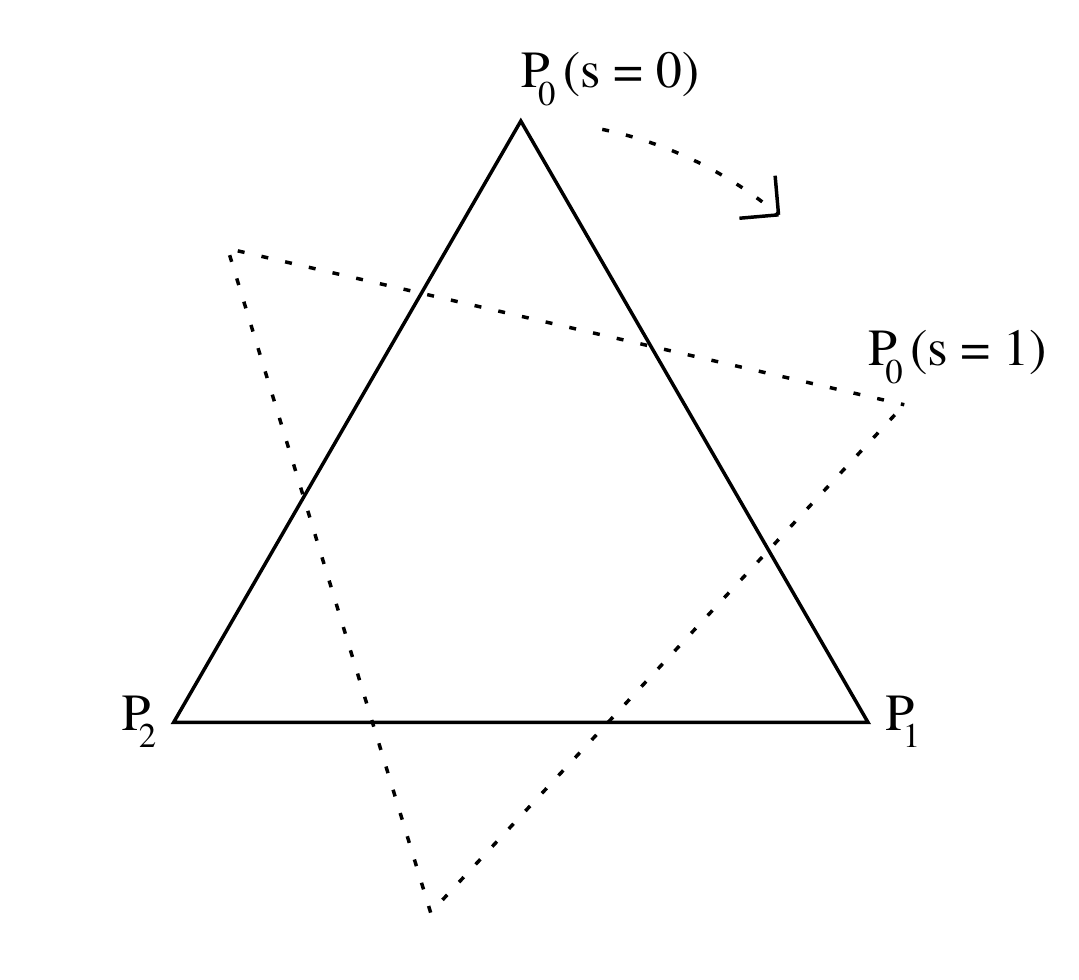}
\caption{An example where the set of instantaneous stationary states forms a simplex, here a triangle.
The extreme points represent the spectral projections $P_i(s),\,i=0,1,2$.
Parallel transport rotates the triangle at time $s=0$ (triangle whose boundary is the full line) to
the triangle at time $s=1$ (triangle whose boundary is the dashed line) as a
rigid body.}
\label{fig:tr}
\end{figure}

Parallel transport may be described more concretely within a particular 
context, as we will show in the next section. For instance, when vectors represent quantum states, the 
instantaneous stationary states are transported like points of a rigid body (see Fig.~\ref{fig:tr}). 

We consider both the case where $\ker L(s)$ is protected by a gap condition, i.e. $0$ is an isolated eigenvalue of $L(s)$, and where it is not. 

\begin{figure}[hbt] 
\begin{center}
\includegraphics[width=13 cm]{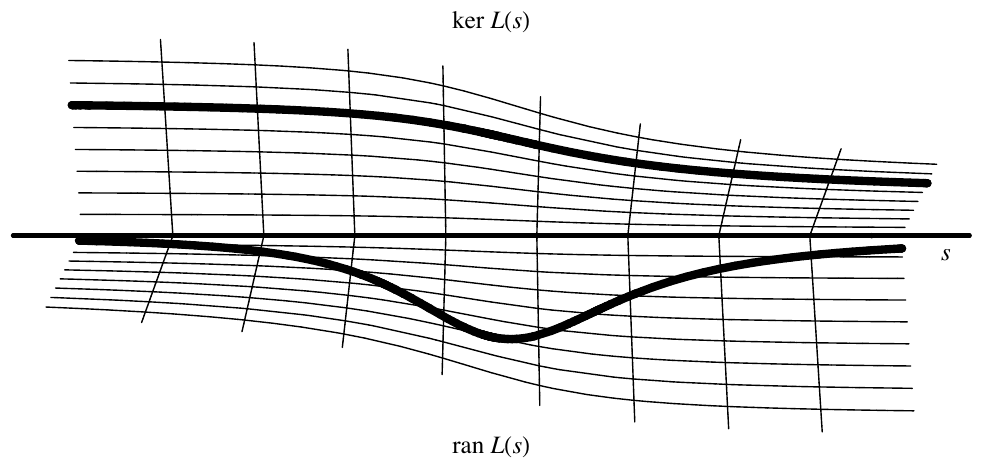}
\caption{The straight lines of the upper and lower bundles represent the kernel and the range of $L(s)$, as they change with $s$ and attain (left and right) asymptotes when $L(s)$ does. The thin curves are the result of parallel transport. The thick curves illustrate the motion within the two subspaces, as  described in Theorem~\ref{thwg} and Corollary~\ref{thm:cor}. It shows the transient nature of the motion in the range: That part is smaller than any power of $\varepsilon$, when $L(s)$ does not vary.}
\label{fig:ker}
\end{center}
\end{figure}

In the gapped case we give an adiabatic expansion which reveals that the 
dynamics has distinct characters within the evolving subspace of 
instantaneous stationary states and transversal to it.
Notably, as we shall see, the motion within $\ker L(s)$ is 
{\it persistent} and partly even {\it irreversible},
whereas the motion transversal to it is {\it transient} in the following
sense: Consider the adiabatic 
evolution over a finite interval, traversed at a slow rate $\varepsilon$;
assume that the generator is constant near
its endpoints and smooth otherwise, and let the initial state be
stationary. Then the distance of the final state from the manifold of
stationary states is exceedingly small in $\varepsilon$
(in fact of infinite order: $O(\varepsilon^N)$ for all $N$), whereas the distance covered within the
manifold is typically $O(1)$ and consists in turn of two parts: A geometric
and potentially reversible part, due to parallel transport, and a subleading {\it irreversible} correction as
large as $O(\varepsilon)$ (see Fig.~\ref{fig:ker}). 
As we will see by the end of the section, 
this single result entails
contrasting physical consequences for isolated and open quantum systems. 

In the gapless case we no longer obtain an expansion, however we prove that the dynamics of the system is
 constrained to the manifold of instantaneous stationary states and is parallel transported along with the manifold as $\varepsilon \rightarrow 0$.
As an application, it generalizes the adiabatic theorem without a gap
condition for the Hamiltonian case (\cite{Bo, AE99, Te01}) to a class of open quantum systems. 
 
Although the framework and the theorems are general and independent of the
context, the geometric interpretation and the implications of the theorems may
depend on it. 
It is instructive to illustrate this point for quantum adiabatic theorems. The
most familiar version is formulated for the {\it Schr\"odinger equation},
where the state of the system is described by a {\it vector} in Hilbert space, and $x$ of Eq.~(\ref{adeq}) is $ \ket{\psi}$. An alternate description could have been given for the {\it von Neumann equation}, where the state
of the system is described by a {\it density matrix}, and $x$ is $\rho$, a positive
matrix with unit trace. For an isolated system, undergoing unitary evolution,
the two descriptions are, in principle, equivalent up to the loss
of an overall phase information in $\rho$; nevertheless the elementary
formulation and direct proofs of the ``standard'' adiabatic theorem tend to
refer to the Schr\"odinger context. A formulation in terms of the density
matrix is of course a prerequisite towards the formulation of adiabatic
theorems for open quantum systems, where only the von Neumann context survives. 
The unified approach presented here gives such a formulation. When applied to the Schr\"odinger equation, it has a precursor in \cite{B90}; when applied to the von Neumann equation, in \cite{Ne93}.

It pays to examine a parallel formulation of the adiabatic theorem for state vectors and density matrices in a simple setting. Consider an isolated, finite dimensional quantum system whose evolution is generated by a slowly varying self-adjoint and non-degenerate Hamiltonian $H(s)$. 
In the context of pure states, where $x= \ket{\psi}$, the manifold of
instantaneous stationary states are the eigenvectors 
associated to a distinguished eigenvalue $e(s)$ and lie in the kernel of
$L(s)=\im\bigl(H(s)-e(s)\bigr)$. Note that $L(s)$ ``knows about'' the
(instantaneous) eigenvalue $e(s)$. The adiabatic theorem then says that the
evolution within the spectral subspace is persistent and depending on
history; to lowest order, it is geometric and encapsulated in Berry's phase \cite{Be}. The evolution transversal to this manifold is transient and non-geometric and describes {\it tunneling} to eigenvectors of different 
eigenvalues.

In the context of density matrices, $x=\rho$, the generator of adiabatic
evolution $L(s)$ is the adjoint action of $H(s)$, namely $L(s)\rho(s)=-\im[H(s),\rho(s)]$. This generator, being invariant to the
replacement $H(s)\to H(s)-e(s)$, has no information on the distinguished
spectral subspace of $H(s)$. Consequently, the manifold of instantaneous
stationary states (in the simplest setting we consider) is a simplex whose
extreme points are the (instantaneous) spectral projections, see Fig.~\ref{fig:tr}. 
In this picture, Berry's phase gets lost; however, the associated curvature remains hidden in the motion transversal to the manifold, as revealed in some instances of linear response theory, like for the quantum Hall effect (see also \cite{afgk}).

We finally consider a class of open quantum systems which, though not
Hamiltonian, preserve the Hamiltonian. The generator of the dynamics, called a
dephasing Lindbladian, retains the above simplex as its manifold of
instantaneous stationary states. If the initial data start at a vertex, the
motion within the manifold of stationary states simply follows the
parametrically rotating vertex (see Fig.~\ref{fig:tr})---this being the
geometric part arising at lowest order---; but to next order the motion is
irreversible, non-geometric and directed away from the vertex. It is
interpreted as {\it tunneling}, in the sense of quantum transitions
between states protected by an energy gap, which may but need not be, a
coherent process.

The anticipated, contrasting consequences are now evident. By the general
result stated earlier, and under its conditions, for systems undergoing
unitary evolution tunneling is {\it reversible}, since it eventually dwindles
to a remainder of infinite order, while for systems governed by a
dephasing Lindbladian, tunneling is {\it irreversible} and comparatively large, $O(\varepsilon)$. 

The plan of the paper is as follows. In Sect.~\ref{sec:res} we describe the
general adiabatic theorems and the properties of parallel transport. In 
Sect.~\ref{sec:appl} we apply these results to unitary and Lindbladian 
quantum systems, as well as to driven stochastic processes. All proofs, 
except for a few short ones, are assembled in Sect.~\ref{proofs}.

\section{General results}\label{sec:res}

In the general scheme mentioned in the Introduction the state space is
a Banach space and the generators are those of 
contraction semigroups. We shall present two adiabatic theorems which, like
their Hamiltonian counterparts, either rely on a spectral gap \cite{ ASY87,K1}
or forgo it \cite{AE99}. Both depend on the notion of parallel transport. 
Some preliminaries, like the existence of the evolution and the definition of
parallel transport, shall be dealt with first. 

\subsection{Preliminaries}
{\it Propagator.\/} We consider the evolution (\ref{adeq}) with 
time-dependent generators $L(s)$, possibly unbounded, 
and state sufficient conditions for the existence of the propagator
on a Banach space $\mathcal{B}$.

\begin{definition}\label{def:ckFamily}
Operators $L(s)$, ($0\le s\le 1$) on $\mathcal{B}$
are called a $C^k$-family if: 
$L(s)$ are closed operators with a common dense domain $D$ and the function 
$L$, taking values in the Banach space of bounded operators $D\to
\mathcal{B}$, is $k$-times differentiable in $s$. Here $D$ is endowed with the graph norm of $L(s)$ for any fixed $s$.
\end{definition}
	\begin{lem}\label{lem:ev} 
	Let $L(s)$, ($0\le s\le 1$) be a $C^1$-family and, for each $s$,
	the generator of a contraction semigroup on $\mathcal{B}$. Then there 
	exist operators 
	$U_\varepsilon(s,s'): \mathcal{B}\to\mathcal{B},\, (0\le s'\le s\le 1)$ with 
	$U_\varepsilon(s,s')D\subset D$, $U_\varepsilon(s,s)=\id$ and
	\begin{equation}
	\label{eq:ev1}
	\varepsilon\frac{\partial}{\partial s}U_\varepsilon(s,s')x
	=L(s)U_\varepsilon(s,s')x,\qquad (x\in D).
	\end{equation}
	For $x\in D$ the unique solution $x(s)\in D$ of (\ref{adeq}) with initial data
	$x(s')=x$ is $x(s)=U_\varepsilon(s,s')x$. Moreover,
	\begin{equation}
	\label{eq:ev1a} 
	\|U_\varepsilon(s,s')\|\le 1 
	\end{equation}
	and
	\begin{equation}
	\label{eq:ev2}
	\varepsilon\frac{\partial}{\partial s'}U_\varepsilon(s,s')x
	=-U_\varepsilon(s,s')L(s')x,\qquad (x\in D).
	\end{equation}
	\end{lem}
	We will call $U_\varepsilon(s,s')x$ a solution of (\ref{adeq}) even for 
	$x\notin D$.
	\begin{rem}
	By definition (\cite{RS2}, Sect. X.8), a contraction semigroup is strongly
	continuous. Its generator is thus closed and densely defined. 
	\end{rem}
	\begin{rem}\label{rem:bounded}
	Suppose, in alternative to the hypothesis of the lemma, that the generator $L(s)$ is bounded and strongly continuous, and that $\varepsilon=1$. Then the propagator exists and is bounded (but not necessarily by 1), uniformly in $0\le s', s\le 1$ \cite{Kr}.
	\end{rem}
\noindent
{\it Parallel transport.\/}
In adiabatic evolutions the manifold of instantaneous stationary states associated to $\ker L(s)$ plays a distinguished role.  This motivates our interest in families of projections. We consider
$P(s):\mathcal{B}\to\mathcal{B}$, ($0\le s\le 1$) to be any 
 $C^1$-family of 
projections in norm sense. Let $\dot{P}(s)=dP(s)/ds$. 
Then the parallel transport $T(s,s'):\mathcal{B}\to \mathcal{B}$ is 
defined by
\begin{align}
 \frac{\partial}{\partial s}T(s,s')&= [\dot{P}(s),P(s)]T(s,s'),
\label{partra} \\
T(s',s') &= \id.  \nonumber
\end{align}
It satisfies $P(s)T(s,s')=T(s,s')P(s')$ and hence respects the ranges of
$P(s)$. Parallel transport is thus a perfect adiabatic evolution: no
transitions from the bundle of projections $P(s)$ to that of the 
complementary projections $Q(s)=\id-P(s)$, nor vice versa.

A characterization of parallel transport, given in terms of sections 
$x(s)\in \ran P(s)$, states that the projected velocity vanishes: 
\begin{equation}
\label{eq:pt}
x(s)=T(s,0)x(0)\; \Leftrightarrow\;P(s)\dot{x}(s)=0, 
\end{equation}
and likewise for $Q$ in place of $P$.
Indeed, for such sections
$\dot{x}=\dot{P}x+P\dot{x}$ and Eq.~(\ref{partra}) reduces to
\begin{equation}
\label{eq:pt2}
\frac{\partial}{\partial s}T(s,s')x(s')=\dot{P}(s)T(s,s')x(s')
\end{equation}
by $P\dot{P}P=0$; hence the contention Eq.~(\ref{eq:pt}). 

The parallel transport determined by the dual projections $P(s)^*:\mathcal{B}^*\to\mathcal{B}^*$ is
\begin{equation}\label{eq:dual}
T^*(s,s')=(T(s',s))^*,
\end{equation}
as can be seen from Eq.~(\ref{partra}). Observe that unless  $\mathcal{B}$ and $\mathcal{B}^*$ coincide, the notion of orthogonal projection does not make sense a priori. In the applications both projections, orthogonal and otherwise, play a role.

It is often the case that open systems evolve towards a unique equilibrium state or a steady state. This situation is associated with rank 1 projections with special properties (see Lemma~\ref{lem:1dimS} and Example~\ref{projection} below) and motivates the interest in this class.

\begin{lem}
\label{lem:1dim}
Let $P(s)$ be a $C^1$-family of rank 1 projections. If $\ker P(s)$ is
independent of $s$, then $\dot{P}(s)$ vanishes on $\ker P(s)$ and 
$P(s)=T(s,s')P(s')$.
\end{lem}

Note that, without making additional assumptions (e.g. that  $P(s)$ is an orthogonal projection),  parallel transport is {\em not} guaranteed to be a contraction. By Remark ~\ref{rem:bounded}, one can only conclude that
\begin{equation}\label{eq:trbound}
\sup_{0\le s', s\le 1}\|T(s,s')\|<\infty.
\end{equation}

\subsection{States} \label{subsec:states}
States of a physical system often enjoy more properties than mere vectors in a
Banach space. The additional structure we will introduce allows for further 
geometric properties of parallel transport. To mark the difference with the
previous and the following subsection, we shall denote states by $\rho$, rather
than by $x$. The fundamental objects, however, are the observables, denoted by $a$, and their algebra $\mathcal{A}$.

In the following let $\mathcal{B}=\mathcal{A}^*$
be the dual of a $C^*$-algebra with
identity $\mathcal{A}$. We consider a second $C^*$-algebra 
$\widetilde{\mathcal{A}}$ and bounded linear maps 
$\Phi: \mathcal{A}\to\widetilde{\mathcal{A}}$ enjoying
\begin{itemize}
\item[(i)] $\Phi$ is positive ($\Phi\ge 0$): $a\ge 0 \Rightarrow \Phi a\ge 0$;
\item[(ii)] $\Phi$ is normalized: $\Phi(\id) = \id$.
\end{itemize}
The maps satisfy $\|\Phi\|=1$ 
(\cite{BR}, Cor. 3.2.6) and form a norm closed 
convex set.
For $\widetilde{\mathcal{A}}=\mathbb{C}$ one is considering linear
functionals, denoted $\rho \in \mathcal{A}^*$, and (i, ii) define {\it states}, 
$\rho \in \mathcal{A}^*_{+,1}$. 
(The subscripts indicate that the functionals
are positive and normalized.) For $\widetilde{\mathcal{A}}=\mathcal{A}$, the
dual maps $\Phi^*:\mathcal{A}^*\to\mathcal{A}^*$ satisfy the corresponding
properties 
(i) $\rho\ge 0 \Rightarrow \Phi^*\rho\ge 0$ and (ii) 
$(\Phi^*\rho)(\id) =\rho(\id)$. We call them {\it state preserving} maps. By
duality,
\begin{equation}
\label{eq:kad}
\|\Phi^*\|=1\,.
\end{equation}
The maps $\Phi$ and $\Phi^*$ then refer to the Heisenberg and the 
Schr\"odinger picture, 
(with $\Phi$ acting on observables and $\Phi^*$ acting on states). We will consider state preserving maps which are 
projections 
$\mathcal{P}: \mathcal{A}^*\to\mathcal{A}^*$. (For economy of notation 
we omit the
star and write $\mathcal{P}_*$ for the predual, if need arises.) 
Associated to them 
are the states in their ranges,
${\mathcal S}:=\mathcal{A}^*_{+,1}\cap \ran \mathcal{P}$. 
Such projections naturally arise through the mean ergodic theorem (\cite{hp},
Thm. 18.6.1) as projections on stationary states of state preserving semigroups $\Phi_t^*$,
\begin{equation*}
\mathcal{P}=
\lim_{\gamma\downarrow 0}\gamma\int_0^\infty \e^{- \gamma t}\Phi_t^*\, d t,
\end{equation*}
provided the limit exists in norm.
\begin{rem}
In case $\mathcal{A}$ does not have an identity, we obtain
$\widehat{\mathcal{A}}$ by adjoining one (\cite{BR}, Def. 2.1.6). We 
consider maps defined on $\widehat{\mathcal{A}}$ satisfying (i, ii), provided
they are compatible with the adjunction. More precisely, we  
consider linear functionals $\rho\in\widehat{\mathcal{A}}^*$ 
(and in particular, {\it states}), provided they arise by canonical extension
from $\rho\in\mathcal{A}$ (\cite{BR}, p. 52). Of a {\it state
preserving} map it is then required to be so also w.r.t. the amended sense of states. 
\end{rem}

	\begin{exa}\label{exa:proj}
	The compact operators $\mathcal{A}=\com(\mathcal{H})$ on a Hilbert space $\mathcal{H}$ 
	form a 
	$\mathcal{C}^*$-algebra (an identity may be adjoined). Its dual $\mathcal{A}^*=\mathcal{J}_1(\mathcal{H})$ are the trace class operators . Any 
	$\rho \in \mathcal{J}_1(\mathcal{H})$ is a state
	if $\rho\ge 0 $ and $\tr \rho = 1$. An example of a state preserving projection 
	$\mathcal{P}$ is 
	\begin{equation}\label{eq:sproj}
	\mathcal{P}\rho=\sum_iP_i \rho P_i, 
	\end{equation}
	where the $P_i$ are an orthogonal partition of unity on
	$\mathcal{H}$. 
	As required by the definition of state preserving maps, $\mathcal{P}$
is the dual of a positive normalized map $\mathcal{P}_*$ on 
	$\com(\mathcal{H})$. In fact, $\mathcal{P}_*$ also acts by (\ref{eq:sproj}).
	\end{exa}

The following proposition is concerned with families of projections 
$\mathcal{P}(s)$ and, more precisely, with the corresponding 
parallel transport $\mathcal{T}(s,s')$, determined by Eq.~(\ref{partra}), and states ${\mathcal S}(s)$: The action of the former on the latter
is that of a rigid motion. In the context of 
Example~\ref{exa:proj} the proposition is illustrated in Fig.~\ref{fig:tr}.
	\begin{prop}[Rigid transport]
	\label{lem:NeumannTr}
	Let the $C^1$-family of projections 
	$\mathcal{P}(s):\mathcal{A}^*\to\mathcal{A}^*$, 
	($0\le s\le 1$) be state preserving. Then 
	$\mathcal{T}(s,s')\mathcal{P}(s')$ 
	is also state preserving.
	In particular, $\mathcal{T}(s,s')$ maps
	\begin{itemize}
	\item ${\mathcal S}(s')$ to ${\mathcal S}(s)$ isometrically;
	\item (isolated) extreme points of 
	${\mathcal S}(s')$ to corresponding 
	ones of ${\mathcal S}(s)$. 
	\end{itemize}
	Moreover, if $\rho(s)\in \ran \mathcal{P}(s)$, depending continuously on $s$, is an 
	isolated extreme point of ${\mathcal S}(s)$, then $\rho(s)=\mathcal{T}(s,s')\rho(s')$.
	\end{prop} 
\noindent
\begin{exa}
(continuing Example~\ref{exa:proj}).
With respect to the
Hilbert-Schmidt inner product induced by the inclusion $\mathcal{J}_1(\mathcal{H})\subset
\mathcal{J}_2(\mathcal{H})$, the projection $\mathcal{P}$ is orthogonal and
the transport $\mathcal{T}$ unitary. These two properties are seen from the
following consideration: Since 
$\mathcal{J}_1(\mathcal{H})\subset\com(\mathcal{H})$, the action of 
$\mathcal{P}_*: \com(\mathcal{H})\to\com(\mathcal{H})$ can be compared with
that of $\mathcal{P}$: $\mathcal{P}_*$ preserves $\mathcal{J}_1(\mathcal{H})$
and $\mathcal{P}_*\restriction\mathcal{J}_1(\mathcal{H})=\mathcal{P}$. We thus
have $\mathcal{T}_*(s,s')\restriction\mathcal{J}_1(\mathcal{H})
=\mathcal{T}(s,s')$ by (\ref{partra}), besides of
$(\mathcal{T}(s,s')\rho)(\mathcal{T}_*(s,s')a)=\rho(a)$ by (\ref{eq:dual}).
\end{exa}

The example is illustrated in Fig.~\ref{fig:tr}. The motion is rigid in the metrics of both
$\mathcal{J}_1(\mathcal{H})$ and $\mathcal{J}_2(\mathcal{H})$. Explicitly, if
$\rho(s)=\sum \lambda_j P_j(s)$ then $\rho(s')=\sum \lambda_j P_j(s')$ for the
same $\lambda_j$, while the projections retain their distances in both norms.

We conclude with a consideration about rank 1 projections, which is linked to
Lemma~\ref{lem:1dim}. In the present setting its hypothesis is satisfied:
	\begin{lem}
	\label{lem:1dimS}
	Consider state preserving projections $\mathcal{P}$ of rank 1. Then 
	$\ran \mathcal{P}_*=\lsp\{\id\}$ and $\ker \mathcal{P}$ is
	independent of $\mathcal{P}$. In particular, if 
	$\mathcal{P}(s)$ is a $C^1$-family of such projections, then
	$\mathcal{P}(s)=\mathcal{T}(s,s')\mathcal{P}(s')$ and 
	$\rho(s)=\mathcal{T}(s,s')\rho(s')$, where $\rho(s)$ is the unique state in 
	$\ran\mathcal{P}(s)$.
	\comment{
	Let $\mathcal{P}(s)$ be a $C^1$-family of rank 1 state preserving projections.
	Then $\ran \mathcal{P}_*(s)=\lsp\{\id\}$ and $\ker \mathcal{P}(s)$ is
	independent of $s$. In particular, 
	$\mathcal{P}(s)=\mathcal{T}(s,s')\mathcal{P}(s')$ and 
	$\rho(s)=\mathcal{T}(s,s')\rho(s')$, where $\rho(s)$ is the unique state in 
	$\ran\mathcal{P}(s)$.}
	\end{lem}
\begin{exa}
Let $\mathcal{A}=\com(\mathcal{H})$ and let 
$\rho_0\in\mathcal{J}_1(\mathcal{H})$ be a state. Then the rank 1 projection 
$\mathcal{P}: \rho\mapsto (\tr \rho)\rho_0$ is state preserving with 
$\ker \mathcal{P} = \{\rho \mid \tr \rho = 0\}$, and $\mathcal{P}_*: a\mapsto
\tr(\rho_0 a)\id $. If $\rho_0=\rho_0(s)$ is a $C^1$-family,
then $\dot{\mathcal{P}}(s)\rho=(\tr \rho)\dot{\rho}_0(s)$ and the statements of the lemma 
are evident. Note
however that, in contrast to the projection (\ref{eq:sproj}), the actions of
$\mathcal{P}$ and $\mathcal{P}_*$ are different. Hence $\mathcal{P}$ is not
orthogonal in $\mathcal{J}_2(\mathcal{H})$. 
\label{projection}
\end{exa}

\subsection{An adiabatic theorem in presence of a gap}\label{gap}
\begin{figure}[hbt]
\begin{center}
\includegraphics[width= 8 cm]{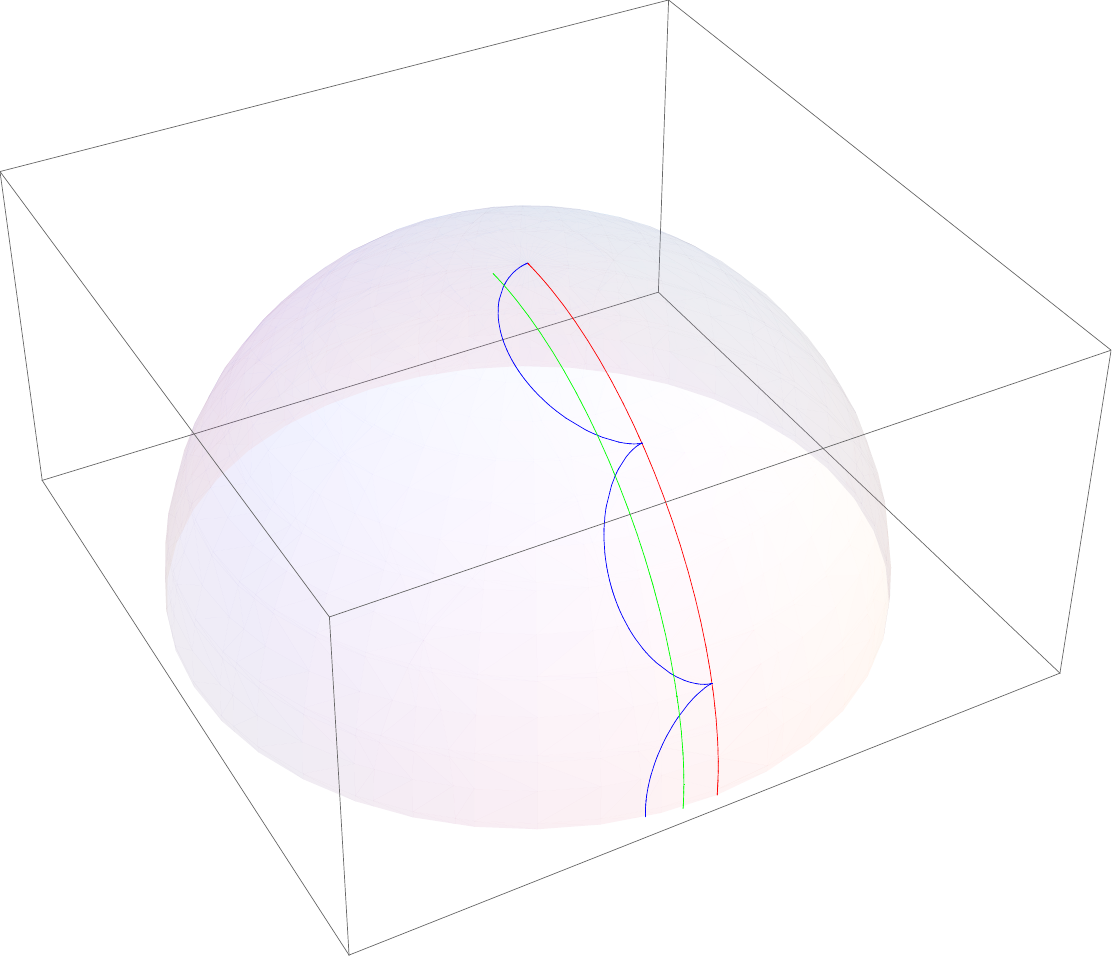}
\caption{
The figure shows the result of a computation of the unitary adiabatic evolution of a qubit, see Subsec.~\ref{subsec:Bloch} for details.  The state is represented as a point on the Bloch sphere, Eq. (\ref{eq:BlochBall}).  The (red)
meridian shows the manifold of instantaneous stationary states,
i.e. $\ker L$. The parametrization corresponds to uniform speed along this path. The ``slow manifold" is represented by the (green) curve essentially parallel to the (red) meridian. An orbit is shown by the (blue) cycloid. Note that the initial conditions do not lie on the slow manifold ($b_1(0)\neq 0$ when $\dot P(0)\neq 0$). This is the reason for the large oscillations.}
\label{fig:bloch2}
\end{center}
\end{figure}

We assume that $0$ is an isolated point of the spectrum
of $L$, which is what we mean by a gap. Then, for small $\varepsilon$, 
the differential 
equation forces a fast time scale of order $O(\varepsilon^{-1})$ on vectors 
transverse to the null space $\ker L(s)$. That scale reflects itself in a fast
motion, consisting of oscillations and decay. By contrast on vectors in the
null space the dynamics is slow by $\dot x=0$. Nevertheless these vectors 
leak out of that subspace, because it is itself changing with $s$.
The leakage however remains of order $O(\varepsilon)$, as shown by
Theorem~\ref{thm:cor2} below. A complementary result, Theorem~\ref{thwg}, 
constructs a ``slow manifold'', where solutions $x(s)$ remain suitably close 
to $\ker L(s)$ and the time scale is $O(1)$. Before presenting the two 
results, which are illustrated in Fig.~\ref{fig:bloch2}, we need to 
specify the transversal subspace complementing $\ker L(s)$.

The general assumptions on $L=L(s)$, ($0\le s\le 1$) are 
\begin{hyp}\label{h1}
$L$ is the generator of a contraction semigroup
on a Banach space $\mathcal{B}$. 
\end{hyp}
As a consequence one has
	\begin{prop} \label{prop:directSum}
	The null space and the range of $L$, the generator of a contraction semigroup, are transversal in the sense that 
	\begin{equation}
	\label{eq:inter0}
	\ker L\cap \ran L=\{0\}.
	\end{equation}
	\end{prop}
The issue whether the two spaces complement each other is covered by the next hypothesis.
  
\begin{hyp}\label{h2}
The range of $L$ is closed and complementary to the (closed)
null space of $L$:
\begin{equation}
\mathcal{B}= \ker L\oplus \ran L, 
\label{eq:tr}
\end{equation}
and the corresponding projections are denoted $\id =P+Q$.
\end{hyp}
\begin{rem}
We recall that $\mathcal{B}_1 \oplus \mathcal{B}_2$ is the
notation for the sum $\mathcal{B}_1 + \mathcal{B}_2$ of subspaces 
$\mathcal{B}_i\subset \mathcal{B}$ ($i=1,2$) in the case that any vector 
$x$ in the sum admits a unique decomposition $x=x_1 + x_2$ with 
$x_i\in\mathcal{B}_i$. 
Any two among the
statements ``$\mathcal{B}_i$ ($i=1,2$) are closed'', 
``$\mathcal{B}_1 + \mathcal{B}_2$ is closed", and 
``$P_i:x\mapsto x_i$ ($i=1,2$) are bounded''
imply the third. 
\end{rem}\begin{hyp}\label{h3} 
$L(s)$ is a $C^k$-family 
for which $0$ remains a uniformly isolated eigenvalue.
\end{hyp}

\begin{rem} 
We will see by 
Hypothesis \ref{h2} that zero is either in the resolvent set or an isolated point of the
spectrum $\sigma(L)$. In the latter case, by 
Hypothesis \ref{h3}, the gap is then assumed 
to be uniform. The restriction $L\restriction\ran L$ has a bounded inverse, 
denoted by $L^{-1}$, and $P(s)$ and $L(s)^{-1}$ are $C^k$ in norm.
\end{rem}
\begin{rem} 
We will give sufficient conditions for 
Hypothesis \ref{h2} in 
Subsec.~\ref{subsec:compl}. For short,  
it is the regular case, given Hypothesis \ref{h1}.
\end{rem}

For $\varepsilon=0$, Eq.~(\ref{adeq}) requires 
$x(s) \in \ker L(s)$. For small $\varepsilon$ the differential 
equation admits solutions which remain close to $\ker L(s)$. 
The construction of the ``slow manifold'' reduces to a differential 
equation for the slow variables only, with the fast ones providing the 
inhomogeneity. The latter, rather than being governed by a further, coupled 
differential equation, are enslaved to the solution at lower orders. More 
precisely, the solutions are described as follows. 
\begin{thm}[Slow manifold expansion] \label{thwg}
Let $L(s)$ be a $C^{N+2}$-family of operators satisfying Hypotheses~\ref{h1}-\ref{h3}. Then
 \begin{enumerate} 
  \item The differential equation $\varepsilon \dot{x}=L(s)x$ admits solutions of the form 
\begin{align} \label{solu}
 x(s)=\sum_{n=0}^N\varepsilon^n(a_n(s)+b_n(s))+\varepsilon^{N+1}r_N(\varepsilon,s)
\end{align}
with
\begin{itemize}
 \item $a_n(s)\in \ker L(s), \ b_n(s) \in \ran L(s)$.
\item initial data $x(0)$ specified by arbitrary $a_n(0) \in \ker L(0), \
r_N(\varepsilon,0) \in \mathcal{B}$; however, the $b_n(0)$ are determined
below by the $a_n(0)$ and together define the "slow manifold".
\end{itemize}
\item The coefficients are determined recursively through $(n=0, \dots, N)$
\begin{align}
 b_0(s)&=0, \nonumber \\
a_n(s)&=T(s,0)a_n(0)+ \int_0^sT(s,s')\dot{P}(s')b_n(s')ds', \label{an} \\
b_{n+1}(s)
&=L(s)^{-1}\dot{P}(s)a_n(s)+L(s)^{-1}Q(s)\dot{b}_n(s). \label{bnpo}
\end{align}
\item The remainder is
\begin{equation}
r_N(\varepsilon,s)=U_\varepsilon(s,0)r_N(\varepsilon,0)+b_{N+1}(s)-U_\varepsilon(s,0)b_{N+1}(0) 
-\int_0^sU_\varepsilon (s,s')\dot{b}_{N+1}(s')ds', \label{rN}
\end{equation}
where $U_\varepsilon (s,s')$ is the propagator described in 
Lemma~\ref{lem:ev}. It
is uniformly bounded in $\varepsilon$, if $r_N(\varepsilon,0)$ is:
\begin{equation*}
\sup_s\|r_N(\varepsilon,s)\|\le 
C_N\sum_{n=0}^N\|a_n(0)\|+\|r_N(\varepsilon,0)\|,
\end{equation*}
where $C_N$ depends on the family.
\end{enumerate}
\end{thm}
Explicitly: for $a_1(0)=0$ we have
\begin{align}
 a_0(s)&=T(s,0)a_0(0), \label{exp1} \\
b_1(s)&=L(s)^{-1}\dot{P}(s)a_0(s), \label{exp2} \\
a_1(s)&=\int_0^sT(s,s')\dot{P}(s')L(s')^{-1}\dot{P}(s')a_0(s')ds'. 
\label{exp3}
\end{align}

\begin{cor}\label{thm:cor}
 If $L(s)$ is constant on an interval $I\subset [0,1]$, then
\begin{align*}
 b_n(s)=0, \qquad (s\in I).
\end{align*}
\end{cor}
\noindent
{\bf Proof.}
This follows recursively from (\ref{bnpo}) by $\dot{P}(s)=0$.
\hfill$\square$

\begin{cor}\label{thm:cor1}
If $P(s)$ are rank 1 projections and $\ker P(s)$ is
independent of $s$, then $a_n(s)=T(s,0)a_n(0)$.
\end{cor}
This is the case in Example~\ref{projection}. See Subsec.~\ref{subsec:thermal} for an application.\\

\noindent
{\bf Proof.} In Eq.~(\ref{an}) we have $\dot{P}(s')b_n(s')=0$ in view of 
Lemma~\ref{lem:1dim} and of 
$b_n(s')\in\ran L=\ker P$.\hfill$\square$\\

In Theorem~\ref{thwg}
	the initial data $x(0)=P(0)x(0)+Q(0)x(0)$ is such that the first (slow) part is arbitrary, and it
	prescribes the second (fast) part, up to a remainder. 
	The general case that both 
	parts of the initial condition are arbitrary is addressed by a result on the decoupling of the slow variables from the fast variables:
	\begin{thm}[Decoupling]\label{thm:cor2} Let $L(s)$ be a $C^2$-family satisfying the assumptions of Theorem \ref{thwg}. 
	Then for any solution $x(s)$ of Eq.~(\ref{adeq})
	\begin{equation*}
	\|P(s)x(s)-T(s,0)P(0)x(0)\|\le C\varepsilon\|x(0)\|,
	\qquad(0\le s \le 1),
	\end{equation*}
	where $C$ depends on the family.\end{thm}
\begin{rem}\label{rem:cor2} No statement about the fast part, $Q(s)x(s)$, is made. The theorem may in particular be
	applied to the difference $\tilde x(0)=Q(0)\tilde x(0)$ of initial conditions 
	sharing the same slow part;
	in this case,
	$\|P(s)\tilde x(s)\|\le C\varepsilon\|\tilde x(0)\|$. 
\end{rem}

\comment{
	In Theorem~\ref{thwg}
	the initial data $\rho(0)=P(0)\rho(0)+Q(0)\rho(0)$ is such that the first (slow) part is arbitrary, and it
	prescribes the second (fast) part, up to a remainder. 
	The general case that both 
	parts of the initial condition are arbitrary is addressed by a result on the decoupling of the slow variables from the fast variables:
	\begin{thm}\label{thm:cor2} Let $L(s)$ be a $C^2$-family satisfying the assumptions of Theorem \ref{thwg}. 
	Then for any solution $x(t)$ of Eq.~(\ref{adeq})
	\begin{equation*}
	\|P(s)\rho(s)-T(s,0)P(0)\rho(0)\|\le C\varepsilon\|\rho(0)\|\,, \qquad (0\le s \le 1)\,,
	\end{equation*}
	where $C$ depends on the family.\end{thm}
	Note that no statement
	about the fast part, $Q(s)\rho(s)$, is made. 
 	The proposition may in particular be
	applied to the difference $\tilde \rho(0)=Q(0)\tilde \rho(0)$ of initial conditions 
	sharing the same slow part;
	in this case,
	$\|P(s)\tilde \rho(s)\|\le C\varepsilon\|\tilde \rho(0)\|$. 
}

	The proof of Theorem~\ref{thm:cor2} will depend on the following result. 
	We consider linear forms $\varphi \in \mathcal{B}^*$, the dual of 
	$\mathcal{B}$. The duality bracket is $\langle \varphi, x \rangle$.
	\begin{prop}[Adiabatic invariants]\label{lm:adinv}
	Let $L(s)$ be a $C^1$-family as above. 
	Suppose the family $\varphi(s) \in \mathcal{B}^*$, $(0\le s\le 1)$ 
	satisfies
	\begin{align}
 	\varphi(s)\in \ker L^*(s), \qquad \dot{\varphi}(s) \in \ran L^*(s). 
	\label{phicond}
	\end{align}
	Then $\varphi$ is an approximate adiabatic invariant in the sense that for any solution $x(t)$ of Eq.~(\ref{adeq})
	\begin{align}
 	\langle \varphi, x \rangle |_0^s = \varepsilon \int_0^s \langle {L^*}^{-1}\dot{\varphi},\dot{x}\rangle ds'. \label{adinv}
	\end{align}
	Assuming $C^2$-regularity, the expression is bounded as
	\begin{align}
	| \langle \varphi, x \rangle |_0^s| \le C\varepsilon\|\varphi(s)\|\|x(0)\|,
	\label{adinv1}
	\end{align}
	where $C$ depends on the family $L(s)$.
	\end{prop}

\subsection{An adiabatic theorem in absence of a gap}\label{nogap}
In the absence of a gap a weaker replacement for the previous theorem 
is provided by the following result, which relies on Hypothesis \ref{h1} and replaces Hypotheses \ref{h2}-\ref{h3} by 
\setcounter{hypprime}{1}
\begin{hypprime}{\hspace{-8pt}'}\label{h2prime}
\begin{equation}
\mathcal{B}= \ker L\oplus \overline{\ran L}.
\label{eq:tr2}
\end{equation}
\end{hypprime}
\begin{hypprime}{\hspace{-8pt}'}\label{h3prime}
$L(s)$ is a $C^1$-family.
\end{hypprime} 
\begin{thm}[Gapless]\label{thwg2}
Let $L(s)$ satisfy Hypotheses \ref{h1}, \ref{h2prime}', and \ref{h3prime}' for all $0\le s\le 1$ and let $P(s)$,  for almost all $s$, be the projection
associated to $\ker L(s)$ in the decomposition (\ref{eq:tr2}); moreover let $P(s)$ be defined for all
$0\le s\le 1$ and $C^1$ as a bounded operator on $\mathcal{B}$. Then the solution of $\varepsilon \dot{x}=L(s)x$
with initial data $x(0)=P(0)x(0)$ satisfies
\begin{equation}
\sup_{s\in [0,1]} \|x(s)-T(s,0)x(0)\|\to 0,\qquad (\varepsilon \to 0)\,.
\end{equation}
\end{thm}

\begin{rem}
The theorem generalizes the Hamiltonian adiabatic theorem in
absence of gap \cite{Bo, AE99} to the
non-self-adjoint case. Actually, the $C^2$-regularity of $P(s)$ assumed there is relaxed here to 
$C^1$ thanks to a remark by Elgart, reported
in \cite{Te01}.
\end{rem}
\begin{rem}\label{rem:aa}
The ``almost all'' formulation \cite{Bo, Te01} allows 
for eigenvalue crossings. 
\end{rem}

Proposition \ref{lm:adinv} has the following variant in the gapless case.

	\begin{prop} \label{lm:adinv1}	
	Let $L(s)$ be a $C^1$-family satisfying Hypothesis 1.
	Suppose the family $\varphi(s) \in \mathcal{B}^*$, $(0\le s\le 1)$ 
	satisfies
	\begin{align}
 	\varphi(s)\in \ker L^*(s), \qquad  \dot{\varphi}(s)=L^*(s)\phi(s) 
	\label{phicond2}
	\end{align}
with uniformly bounded $\phi(s)$ and $\dot{\phi}(s)$.  Then 
	\begin{align}
	| \langle \varphi, x \rangle |_0^s| \le 
        3\varepsilon \sup_{0\le s'\le 1}
\bigl(\|\phi(s')\|+\|\dot{\phi}(s')\|\bigr)\|x(0)\|.
	\label{adinv2}
	\end{align}
	\end{prop}

\subsection{Complementarity of subspaces}\label{subsec:compl}
In this subsection we will give sufficient conditions for the complementarity 
Hypothesis \ref{h2} 
in relation with a spectral gap, and 
2' in its absence.
As a help to gauge them, note that 
both are false if $0$ is an
eigenvalue of $L$ with non-vanishing eigennilpotent, but they hold true for a
self-adjoint operator $L$ on a Hilbert space if $0$ is an isolated
resp. non-isolated point of its spectrum.   

The two subspaces in Eqs.~(\ref{eq:tr}, \ref{eq:tr2}) are transversal,
\begin{equation*}
\ker L\cap \overline{\ran L}=\{0\},
\end{equation*}
as a consequence of Hypothesis \ref{h1}, as we shall see.
However they may fail to generate $\mathcal{B}$ without further hypotheses. Such hypotheses are given in two lemmas corresponding 
to the two cases. There, a prime indicates a hypothesis tailored to the
second, gapless case; a sufficient condition for an earlier hypothesis is noted by an added roman numeral.

Counterexamples matching the two cases are also
given. Related results are found in (\cite{hp}, Thm. 18.8.3).

\begin{lem}\label{lem:tr}
Let $\mathcal{B}$ be a Banach space and $L$ a closed operator on
$\mathcal{B}$. Assume, besides Hypothesis \ref{h1}, that
\begin{itemize}
\item[(H2i)] If $0$ is in the spectrum, $\sigma(L)$, then $0$ is a discrete eigenvalue.
\end{itemize}
Then $\mathcal{B}= \ker L\oplus \ran L$, cf. Eq.~(\ref{eq:tr}).

Property (H2i) implies that $L$ is Fredholm, and hence 
\begin{itemize}
\item[(H2ii)] $L$ is semi-Fredholm. 
\end{itemize}
In conjunction with
Hypothesis \ref{h1},  
Properties (H2i) and (H2ii) are
equivalent.
\end{lem}
\noindent
\begin{rem}
  By definition, a discrete eigenvalue is an isolated 
point $\lambda$ of the spectrum, the Riesz projection of which,
\begin{equation}
\label{eq:rp}
P=-\frac{1}{2\pi \im}\oint_\Gamma(L-z)^{-1}dz,
\end{equation}
is finite-dimensional. Here $\lambda$ is the only point of the spectrum
encircled by $\Gamma$.
\end{rem}
\begin{rem} 
Hypothesis (H2i) is trivially satisfied if
$\dim\mathcal{B}<\infty$. 
\end{rem}
\begin{rem} 
We recall that $L$ is semi-Fredholm iff $\ran L$ is
closed and $\ker L$ or $\mathcal{B}/\ran L$ are finite-dimensional. If both
are, $L$ is called Fredholm.
\end{rem}

\begin{exa}
Assumption (H2i) can not be omitted from Lemma~\ref{lem:tr} if
the splitting (\ref{eq:tr}) is to be ensured. 
In fact in (\cite{LuPhi61}, Thm. 2.2) an example is given of a non-trivial
generator $L$ of contraction semigroup with trivial null space, yet with 
$\sigma(L)=\{0\}$. Hence (H2i) fails there. By the
equivalence with (H2ii), $\ran L$ is not closed, spoiling (\ref{eq:tr}).
\end{exa}
\begin{lem}\label{lem:tr2}
Let $\mathcal{B}$ be a Banach space and $L$ a closed operator on
$\mathcal{B}$. Assume, besides Hypothesis~\ref{h1}, that 
\begin{itemize}
\item[(H2'i)] $\mathcal{B}= \ker L+ \overline{\ran L}$.
\end{itemize}
Then $\mathcal{B}= \ker L\oplus \overline{\ran L}$, cf.~Eq.~(\ref{eq:tr2}).

Moreover, if $\ker L+ \overline{\ran L}$ is closed and $\mathcal{B}$ 
reflexive, then 
(H2'i) follows from Hypothesis~\ref{h1}. 
\end{lem}

\noindent
Recall that, by definition, $\mathcal{B}$ is reflexive if 
$\mathcal{B}^{**}=\mathcal{B}$.

\noindent
\begin{exa}\label{exgang}
Consider the operator $L$ defined by 
$(Lf)(x)=-xf(x)$ for $f\in L^\infty(0,1)=\mathcal{B}$. Obviously, $L$ has
trivial kernel and $(\e^{Lt}f)(x)=\e^{-xt}f(x)$, which makes $L$ the generator of a
contraction semigroup. 
However, for $1 \equiv g \in L^\infty(0,1)$ 
one has
\begin{align*}
 \|g-L f\|_{L^\infty} \geq 1 \,,\qquad (f \in L^\infty(0,1))\,.
\end{align*}
Thus $\overline{\ran L}$ is a proper subspace of $L^\infty(0,1)$.
In relation with Lemma~\ref{lem:tr2}, the example shows that
when $\mathcal{B}$ is not reflexive (H2'i) does not follow from Hypothesis~\ref{h1}. 
\end{exa}

\begin{exa}\label{exlind} As a further, similar example consider the 
operator $L:\rho\mapsto -\im[H,\rho]$ defined for 
$\rho\in\mathcal{J}_1(\mathcal{H})=\mathcal{B}$, where $H$ is a bounded self-adjoint
operator on the Hilbert space $\mathcal{H}$. Let $H$ have purely continuous
spectrum, so that $\ker L=\{0\}$. On the other hand, $\tr L\tilde\rho=0$ for any $\tilde{\rho} \in \mathcal{J}_{1}(\mathcal{H})$, because $\tr H\tilde\rho=\tr\tilde\rho H$ (\cite{Si79}, Cor. 3.8). Then $\tr\rho=0$ extends to $\rho\in\overline{\ran L}$, which is thus a proper subspace of 
$\mathcal{J}_1(\mathcal{H})$.
\end{exa}

\section{Applications}\label{sec:appl} 
Our results apply to a wide range of driven quantum and classical
systems.
For quantum systems we consider evolutions generated either by a Hamiltonian
or a Lindbladian. We focus on the special class of ``dephasing Lindbladians"
which are in some sense intermediate between Hamiltonians and generic 
Lindbladians. As we shall explain, adiabatic evolutions in the Hamiltonian
setting have a different character from those in the dephasing setting. In
the Hamiltonian case tunneling is reversible while in the dephasing one it is irreversible.

In classical systems we consider continuous-time Markov processes. We give 
an adiabatic expansion for a slowly driven Markov process with unique 
stationary distribution and then restrict our attention to reversible 
processes and to the generation of probability currents.

\subsection{Unitary evolutions}\label{subsec:uev}
The results of Sect.~\ref{sec:res} may be applied to recover known facts about the unitary adiabatic evolution 
driven by smoothly varying family of self-adjoint Hamiltonians $H(s)$ on a
Hilbert space $\mathcal{H}$ \cite{K1, Ne93}. Consider a simple, discrete eigenvalue 
$e(s)$ of $H(s)$. Its normalized eigenfunction $\psi(s)$ spans the manifold 
of instantaneous stationary states, i.e. the kernel of 
\begin{align}\label{schrod}
 L(s)=-\im(H(s)-e(s)).
\end{align}
Eq.~(\ref{adeq}) is the Schr\"odinger equation and let $\psi_\varepsilon(s)$ 
be its solution with initial data $\psi_\varepsilon(0)=\psi(0)$. Tunneling, $T(s)$, is defined as the leaking out from the manifold of stationary states, i.e.
\begin{equation}\label{eq:tun}
T_\varepsilon(s)= 1-|(\psi(s),\psi_\varepsilon(s))|^2.
\end{equation}
There is extensive literature (see \cite{hagedorn} and references therein) which is concerned with estimates 
of the tunneling amplitude at all orders in $\varepsilon$, or beyond. 
 The simplest version of these results can be seen to be a consequence of 
 Corollary~\ref{thm:cor}. Namely:
 \begin{thm} 
 \label{expTunneling}
Suppose that $H(s)=H(s)^*$ is a $C^\infty$-family in the sense of Definition \ref{def:ckFamily}, and 
is in addition constant near the endpoints $s=0$ and $s=1$. Let $\psi_\varepsilon(s)$ and $\psi(s)$ be as above. Then
$T_\varepsilon(1)=O(\varepsilon^k)$, for any $k$, see
Fig.~\ref{fig:ker}.
\end{thm}

It may be instructive to examine this result from the perspective of the
evolution of density matrices generated by the adjoint action of $H$. In
contrast with the Schr\"odinger generator of Eq.~(\ref{schrod}) whose kernel
is one dimensional, the adjoint action $-\im[H(s),\rho]$  has a large kernel
spanned by all the stationary states. Tunneling is then described not by the
motion in the range but rather by the motion in the kernel. Theorem~\ref{expTunneling} may then be interpreted as the statement that the adiabatic map is a rigid map of the kernel up to terms of infinite order, at time one, see Fig. \ref{fig:tr}. 

Interestingly, when Theorem~\ref{thwg} is applied to open quantum systems,
described by a dephasing Lindbladian one reaches the opposite conclusion, namely that tunneling is irreversible and $O(\varepsilon)$.
To complete the picture, it may be worthwhile to discuss from the latter perspective why tunneling gets reduced from first to infinite order in $\varepsilon$ in the special case of the adjoint action. These two points will be addressed in detail in Corollary~\ref{cor:lind} and thereafter.

\subsection{Evolutions generated by Lindblad operators}\label{subsec:thermal}

Lindbladians arise as generators of 
dynamical semigroups \cite{Li76, Davies}. Different settings are available in the
literature. We choose one of them: Let $\mathcal{A}$ be a $C^*$-algebra with
identity and let $\Phi_t$, 
($t\ge 0$) be a norm-continuous semigroup of positive normalized maps
on $\mathcal{A}$. As noted 
in Subsec.~\ref{subsec:states}, $\Phi_t$ and $\Phi_t^*$ have norm $1$, and hence 
are contraction semigroups, with dual generators, $\mathcal{L}_*$ and 
$\mathcal{L}$.

Following \cite{Li76}, $\Phi_t$ is called a {\it dynamical
semigroup} if, besides of the above properties, it is completely positive. We
find it convenient to follow the accepted tradition and call
 the generator in the Schr\"odinger picture, 
$\mathcal{L}=(\mathcal{L_*})^*$, the {\it Lindbladian}.

More precisely, the generator in the Heisenberg picture $\mathcal{L_*}$ is a
(weak-* continuous) operator on the Banach 
space $\mathcal{B}(\mathcal{H})$ of bounded operators on  a Hilbert space
$\mathcal{H}$. It is thus determined by its restriction on 
$\mathcal{A} = \mathrm{Com}(\mathcal{H})$, the compact operators on 
$\mathcal{H}$ (cf. examples of Sect.~\ref{sec:res}).
Then the Lindbladian 
$\mathcal{L}\,:\, \mathcal{J}_1(\mathcal{H}) \to \mathcal{J}_1(\mathcal{H})$
has
a general form \cite{Li76}
\begin{equation}
 \mathcal{L}\rho = -\im[H,\rho]+\frac 1 2 \sum_\alpha\big([\Gamma_\alpha \rho, \Gamma_\alpha^*]+[\Gamma_\alpha, \rho \Gamma_\alpha^*]\big) \label{eq:lin}
\end{equation}
with $H = H^*$ and $\sum_\alpha\Gamma_\alpha^*\Gamma_\alpha$ bounded operators on $\mathcal{H}$. 

\begin{rem}\label{rem:energy}
 $\mathcal{L}$ does not determine $\Gamma_\alpha,\,H$ uniquely. The ``gauge transformation''
$\Gamma_\alpha\to \Gamma_\alpha+ \beta_\alpha \id$ and $H\to H+
\im\sum_\alpha(\beta_\alpha\Gamma_\alpha^*-\overline{\beta_\alpha}\Gamma_\alpha)/2$
leave $\mathcal{L}$ invariant.
 \end{rem}
 
In the generic case the Lindbladian has a $1$-dimensional kernel, with $\ker
\mathcal{L}_*=\lsp\{\id\}$ independently of $\mathcal{L}$, 
cf. Lemma~\ref{lem:1dimS}. 
We consider a smoothly varying family of Lindbladians. 
Let $\rho(s)$ be the corresponding state and 
$\rho_\varepsilon(s)$ be the solution of the adiabatic evolution equation Eq.~(\ref{adeq}) with initial data 
$\rho_\varepsilon(0)=\rho(0)$. The tunneling Eq.~(\ref{eq:tun}) should be 
generalized to $T=1-F^2$, where the fidelity is
\begin{equation}
\label{eq:fidelity}
F_\varepsilon(s)=\tr\bigl((
\rho(s)^{1/2}\rho_\varepsilon(s) \rho(s)^{1/2})^{1/2}\bigr).
\end{equation}

In the presence of a gap a system relaxes to its equilibrium state exponentially fast. A gapped system with a unique ground state will remain close to the instantaneous equilibrium state under adiabatic deformation. For the more interesting, gapless case, see Subsec.~\ref{subsec:gapless}. From the results in the previous section we have:
	\begin{thm}\label{LindTunneling}
	Let $\mathcal{L}(s)$ be $C^\infty$-family of Lindbladians 
	having a unique instantaneous
	stationary state $\rho(s)$. Then, by Lemma \ref{lem:1dim}, $\rho(s)$ is parallel transported. Suppose
	that the assumptions of
	Theorem~\ref{thwg} hold  and is $L(s)$ is constant near the endpoints $s=0, 1$. 
	Then, by Corollary~\ref{thm:cor1}, $\rho_\varepsilon(1)=\rho(1)+O(\varepsilon^k)$ for any $k$ and the tunneling out of the ground state is
	$T_\varepsilon(1)=O(\varepsilon^k)$. 	\end{thm}
\subsection{Dephasing Lindbladians}\label{subsec:Bloch}
 We say that 
$\mathcal{L}$ is a dephasing Lindbladian (corresponding to a given 
Hamiltonian $H$) if 
\begin{equation}\label{eq:dldef}
 \ker \mathcal{L}_*\supset\ker([H,\,\cdot])
\end{equation}
as subspaces of $\mathcal{B}(\mathcal{H})$.

By the following proposition, the evolution
 shares the manifold of stationary states with the corresponding Hamiltonian
evolution. 
	\begin{prop}\label{lm:deph} In connection with Eq.~(\ref{eq:lin}) we have:
	\begin{enumerate} 
	\item 
	$\ker \mathcal{L}_*\supset\ker([H,\,\cdot])$ is equivalent to 
	$\Gamma_\alpha=f_\alpha(H)$ for some functions $f_\alpha$.
	\item 
	$\Gamma_\alpha=f_\alpha(H)$ implies 
	$\ker \mathcal{L}=\ker([H,\,\cdot])$ as subspaces of $\mathcal{J}_1(\mathcal{H})$.
	\item 
	If the spectrum of $H$ is pure point, then the last
	implication is an equivalence. This applies in particular to the finite-dimensional case.
	\end{enumerate} 
	\end{prop}
A dephasing Lindbladian conserves all observables which are conserved
by $H$, in particular $H$ itself
and $\Gamma_\alpha$.  If one interprets the energy of the system in terms of
$H$ and $\Gamma$ (see Remark~\ref{rem:energy}) then one learns that although
the system is open, it does not exchange energy with a bath. However, the dephasing Lindbladian induces decoherence w.r.t. the energy eigenbasis. A (non-rigorous) scenario where that may arise is discussed in \cite{PZ}.
\begin{exa}\label{example:Bloch0}
The simplest dephasing Lindbladian is a 2-level system (a qubit). It is a 4-parameter family: The Hamiltonian is determined by the 3-vector $b$
\begin{equation*}
2 H = b \cdot \sigma, \qquad (b \in \mathbb{R}^3,\,\sigma=(\sigma_1, \sigma_2,\,\sigma_3))\,
\end{equation*}  
where $\sigma_j$ are the Pauli matrices and $\gamma\ge 0$ characterizes the dephasing
\begin{equation}
\label{eq:lin2}
\mathcal{L}\rho = -\im[H,\rho] +\gamma|b|^{-1}\big[ [H,\rho],H\big],\qquad 
(\gamma\ge 0).
\end{equation}
(Recall that by $4H^2= (b\cdot b) \id$ any function of $H$ is of the form 
$f(H)= \alpha H +\beta \id$; the dephasing term is written in such a way that
$\gamma$ is dimensionless.)
The canonical map of normalized states into the Bloch ball, 
\begin{equation}\label{eq:BlochBall}
\rho \mapsto n \in \mathbb{R}^3,|n| \leq 1 : \quad 
\rho = \frac{\id + n\cdot \sigma}{2},
\end{equation}
maps the evolution equation $\dot\rho = \mathcal{L} \rho$ into the 
Bloch equation \cite{gorini}
\begin{equation}\label{eq:bloch}
\dot n = b \times n + \gamma \hat{b} \times ( b \times n),
\end{equation}
where $\hat{b}=b/|b|$.
\end{exa}

\subsection{Adiabatic expansion for dephasing Lindbladians}

For simplicity consider $H$ with simple eigenvalues $e_0, \dots, e_{d-1}$ with
normalized eigenvectors $\psi_i$:
\begin{align*}
 H=\sum_i e_i P_i, \qquad P_i=|\psi_i\rangle \langle \psi_i |
\end{align*}
on a finite dimensional Hilbert space $\mathcal{H}$, $\dim \mathcal{H} = d$.
The operators $E_{ij}:= |\psi_i \rangle \langle \psi_j |$ form a basis of
$\mathcal{B}(\mathcal{H})$, the linear maps on $\mathcal{H}$, which is
orthonormal once that space is endowed with the Hilbert-Schmidt inner product. 
A straightforward computation using Prop.~\ref{lm:deph} shows that $E_{ij}$ are eigenvectors of $\mathcal{L}$ and 
the eigenvalues $\mathcal{L}E_{ij} = \lambda_{ij} E_{ij}$ satisfy 
 $\lambda_{ij}=\overline{\lambda}_{ji},\, \Re \lambda_{ij} \leq 0$ and
$\lambda_{ij} = 0$ if and only if $i = j$.
Hence $\ker \mathcal{L}$ is spanned by $E_{ii}=P_i$ and $\ran \mathcal{L}$ by $E_{ij}$, ($i\neq j$) with the corresponding projections (cf. (\ref{eq:sproj}))
\begin{align*}
 \mathcal{P}\rho=\sum_i P_i \rho P_i, \qquad \mathcal{Q}\rho=\sum_{i \neq j} P_i \rho P_j.
\end{align*}

\begin{figure}[hbt] 
\begin{center}
\includegraphics[height=4 cm]{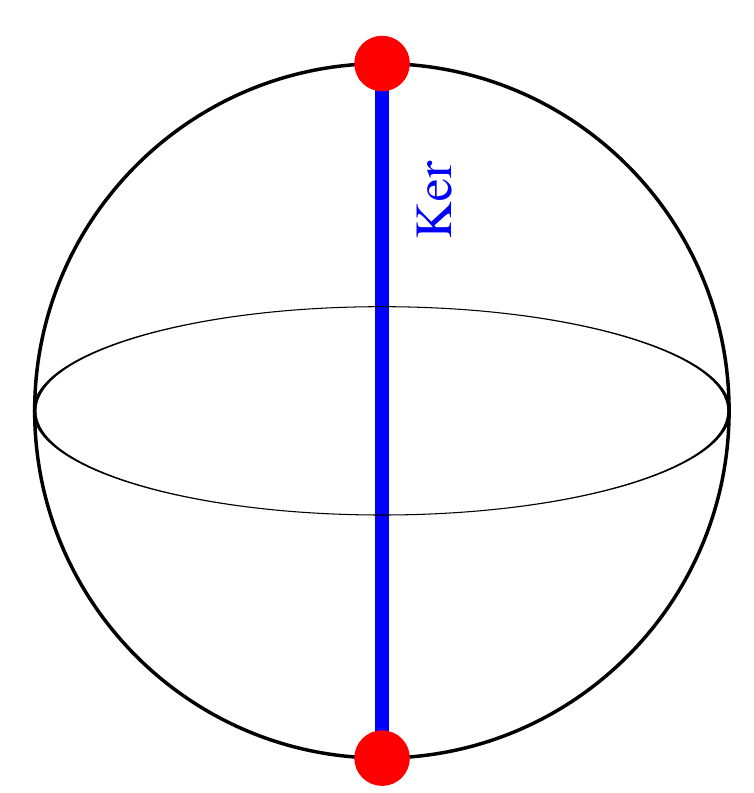}
\caption{The states of a qubit (2-level system) can be represented as the 3D
ball, the interior of the Bloch sphere. For a dephasing Lindbladian, the set
of stationary states is the (blue) axis whose extreme points (red dots) are
spectral projections for the Hamiltonian $H$. In the adiabatic setting
the (blue) axis moves slowly.}
\label{fig:bloch}
\end{center}
\end{figure}

We now consider a smooth family of Lindbladians $\mathcal{L}(s)$ of dephasing form.

	\begin{thm}\label{thm:lind}
        The equation
	\begin{align*}
 	\varepsilon \dot{\rho}(s) = \mathcal{L}(s) \rho(s)
	\end{align*}
	admits a solution of the form
	\begin{align}
 	\rho(s)= P_0(s) +\varepsilon \sum_{j\neq 0} \left(\frac{ P_j
	\dot{P}_0}{\lambda_{j 0}} + \frac{\dot{P}_0 P_j}{\lambda_{0j}}\right)
	-\varepsilon\sum_{j\neq 0}\big(P_0(s)-P_j(s)\big) \int_0^s \alpha_j(s')ds' + O(\varepsilon^2) \label{Lindsol}
	\end{align}
	with
	\begin{align*}
 	\alpha_j(s)=\tr (P_0(s)\dot{P}_j(s)^2 P_0(s))\cdot \frac{(-2 \Re \lambda_{0j}(s))}{|\lambda_{0j}(s)|^2} \geq 0\,.
	\end{align*}
More generally, the expansion applies to any solution for which it does at $s=0$, e.g. for the one with initial condition $\rho(0)= P_0(0)$, if $\dot{P}_0(0) = 0$.
	\end{thm}
The expansion (\ref{Lindsol}) is just $\rho(s)=a_0(s) + \varepsilon
(b_1(s)+a_1(s)) + O(\varepsilon^2)$, in this order, with
coefficients given in (\ref{exp1}--\ref{exp3}). Like in the Hamiltonian case,
$b_1(s) \in \ran \mathcal{Q}(s)$ describes 
the shift of the slow manifold relative to the manifold of instantaneous stationary states which is reversible in the sense of Corollary~\ref{thm:cor}. Unlike there, $a_1(s) \in \ran
\mathcal{P}(s)$ now describes irreversible tunneling by means 
of a loss and a gain term involving $P_0(s)$ and $P_j(s)$, ($j\neq 0$) respectively. More quantitatively, tunneling out of $P_0(s)$ is given by Eq.~(\ref{eq:fidelity}) (with $\rho(s)$ there replaced by the rank 1 projection $P_0(s)$) 
as
$$T_{\varepsilon}(s) =1 - F^2_{\varepsilon}(s)=1 - \tr(\rho(s) P_0(s)).$$
For arbitrary $\dot P_0(0)$ we have the
following result:
\begin{cor}\label{cor:lind}
The solution of 
$\varepsilon \dot \rho(s) = \mathcal{L}(s) \rho(s)$ with the initial condition
$\rho(0) = P_0(0)$ tunnels like
$$
T_\varepsilon (s) = \varepsilon \sum_{j \neq 0} \int_{0}^{s} \alpha_j(s')ds'  + O(\varepsilon^2) 
$$
with $\alpha_j(s)\ge 0$: Tunneling occurs at a non-negative rate, is irreversible and $O(\varepsilon)$.
\end{cor}
This result should be contrasted with the small tunneling of infinite order in the unitary case, Theorem~\ref{expTunneling}. Alternatively, that case can be analyzed on the basis of $\mathcal{L}(s)\rho=-\im[H(s),\rho]$, following \cite{Ne93}. The solution clearly remains a projection from $\rho(0)=P_0(0)$ on, i.e. $\rho(s) = \rho(s)^2$.
Using the expansion (\ref{solu}) for $x(s)=\rho(s)$ then yields
\begin{equation*}
\label{3}
a_n + b_n = \sum_{j =0}^n (a_j a_{n-j} + a_j b_{n-j} + b_j a_{n-j} + b_j b_{n-j})\,.
\end{equation*}
In view of $a_0(s)=P_0(s)$, see Example~\ref{exa:proj}, and $b_0(s)=0$, this reads $a_n=c_n+a_nP_0+P_0a_n$, where $c_n$ depends on $a_0,\ldots, a_{n-1},b_0\ldots,b_n$. Since $\mathcal{P}a_n=a_n$ by definition of $a_n$, we obtain two recursions for $P_ia_nP_i$, one for $i=0$ and one for $i\neq 0$. Together with Eq.~(\ref{bnpo}), all coefficients $a_n(s), b_n(s)$ are now determined instantaneously in terms of $H(s)$ and its derivatives, and in particular without reference to the history $H(s')$, ($s'<s$). As a result, the tunneling is of infinite order.

\comment{ 
\begin{figure}[hbt] 
\begin{center}
\includegraphics[width= 8 cm]{slave.pdf}
\caption{The figure illustrates the adiabatic expansion in Example~\ref{example:Bloch} in the case of no dephasing $\gamma=0$. It shows the northern hemisphere of the Bloch sphere, Eq.~\ref{eq:BlochBall}. The orbit of $H(s)$ is represented by the thin (red) meridian starting at the north pole. The parallel thick (green) curve shows the shift due to the term ``geometric magnetism'' for the case of uniform speed. }
\end{center}
\label{fig:slave}
\end{figure}
}
\begin{exa}\label{example:Bloch}
(continuing Example~\ref{example:Bloch0}) 
The adiabatic expansion Eq.~(\ref{Lindsol}) takes a rather simple form for the
Bloch equation (\ref{eq:bloch}). With $\dot n$ replaced by 
$\varepsilon\dot n$ and initial condition
$n(0) = -\hat b(0)$ one finds
\begin{equation}
n(s) = -\hat b(s) +\frac{\varepsilon}{ |b(s)|}\left(\frac{\gamma(s)\dot{\hat b}(s)
+
\hat b(s) \times \dot {\hat b}(s)}{1 + \gamma^2(s)} 
+ b(s) \int_0^s \alpha(t) dt\right)
+O(\varepsilon^2),
\label{eq:exp2}
\end{equation}
where
$$
\alpha(t)= \frac{\gamma(t)}{1+\gamma^2(t)}\frac{ |\dot {\hat b}(t)|^2}{|b(t)|}.
$$
The terms in parentheses, in the order as they appear, have the following
interpretation: The first term, being proportional to $\gamma \dot{\hat b}(s)$
describes friction that causes lagging behind the driver $\hat b$. The second
term describes ``geometric magnetism", a term introduced by \cite{BR93}. The
third term is tunneling and describes motion along the axis towards the
center, see Fig.~\ref{fig:bloch}.
While the first two terms describe
instantaneous response in the plane perpendicular to the stationary
axis $\hat b(s)$, the last term describes irreversible motion 
inside the Bloch sphere along the axis, Fig.~\ref{fig:bloch}.
\end{exa}
{\bf Proof.} That (\ref{eq:lin2})
 defines the most general dephasing Lindbladian follows from its 
spectral properties, since the Lindbladian is uniquely determined by kernel
and an off-diagonal eigenvalue $\lambda,\,\Re \lambda \leq 0$; if $0$ stands
for the ground state, then $\Im\lambda\geq 0$ for $\lambda=\lambda_{01}$.
The Bloch equation follows from the commutation relations 
$[n_1 \cdot \sigma,\,n_2 \cdot \sigma] = 2 \im (n_1 \times n_2) \cdot \sigma$. 
To get the expansion (\ref{eq:exp2}) write (\ref{Lindsol}) in the form
\begin{multline}
 \rho(s)= P_0(s) +\frac{\varepsilon}{|\lambda|^2} \bigl(\Re \lambda \{P_1,\,
\dot{P}_0\} + \im\,\Im \lambda [P_1,\,\dot{P}_0]\bigr) \nonumber \\ 
-\varepsilon (P_0(s)-P_1(s)) \int_0^s \alpha_1(s')ds' + O(\varepsilon^2)
\nonumber
\end{multline}
and use $\lambda=|b|(\im-\gamma)$ as well as the (anti-)commutation relations
$$
\{P_1,\dot{P}_0\} = -\frac{1}{2} \dot{\hat b} \cdot \sigma, \qquad
[P_1,\dot{P}_0] = -\frac{\im}{2}(\hat{b} \times \dot{\hat b}) \cdot \sigma,
\qquad (\dot P_1)^2 = \frac{|\dot{\hat b}|^2}{4}
$$
to get the first order correction terms exactly in the same order as they
appear in (\ref{eq:exp2}). \linebreak\phantom{x}
\hfill $\square$

Solutions of the Bloch equations are illustrated in Fig.~\ref{fig:bloch2}.

Further applications of driven dephasing Lindbladians are described in \cite{afgg1, afgg2, afgk}.

\subsection{Driven Markov processes}
\label{subsec:nick}
Theorem~\ref{thwg} may be applied to an evolution of the probability 
distribution of a continuous-time Markov process. In particular, we shall  describe below an application to (stochastic) molecular pumps \cite{RHJ08} (see also \cite{Pa98, HJ09}).  

Let $X$ be a random variable on a finite state space 
$S = (1,\,2,\dots,\,d)$ and denote 
$$
p_i = \mathrm{Prob}(X=i). 
$$
The evolution of $X$ is governed by 
\begin{equation}
\dot{p}_i = \sum_{j=1}^d L_{ij} p_j,
\end{equation}
where the transition rate $j \to i$, $L_{ij}\,(i \neq j)$, is non-negative
and $L_{jj} := -\sum_{i\neq j} L_{ij}$.
The transition matrix $\phi(t):=\exp(L t)$ is a left-stochastic matrix 
($0 \leq \phi_{ij} \leq 1$, $\sum_{i=1}^d \phi_{ij} = 1$), a contraction 
in the 
norm $\|p\|_1=\sum_j|p_j|$, and converges to a projection, $\phi(t)\to P^+$,
($t\to\infty$) (\cite{stochastic}, Thm. 4.4.8).
The range of $P^+$ is spanned by stationary probability 
distributions, meaning
$\sum_{j =1}^d L_{ij} \pi_j = 0$. 

We assume that the state space $S$ is indecomposable and denote by $\pi$ 
the unique stationary distribution 
of $L$, whence $\ker L =\lsp\{\pi\}$ and $\ran L=\{p\mid \sum p_i=0\}$. 
In line with Eq.~(\ref{eq:tr}), let $P$ be the rank 1 projection
associated to that pair of subspaces, which are left invariant by $L$. 
We identify $L^{-1}$ 
with the map $(1-P)L^{-1}(1-P)$ defined on all of $\mathbb{C}^d$, and denote 
its matrix elements by $L^{-1}_{ij}$.

Now we consider a smooth family of generators $L(s)$ with corresponding
stationary states $\pi(s)$.

\begin{thm} \label{thm:mar}
Assume that $S$ is indecomposable for $L(s)$ and that $\dot \pi_i(0) = 0$.
The solution of
\begin{equation}
\varepsilon \dot p_i(s) = \sum_{j =1}^d L_{ij}(s) p_j(s) 
\end{equation}
with initial condition $p_i(0) = \pi_i(0)$ is
\begin{equation}
p_i(s) = \pi_i(s) + \varepsilon \sum_{j=1}^d L^{-1}_{ij}(s) \dot \pi_j(s) + O(\varepsilon^2).
\label{eq:expmar}
\end{equation} 
\end{thm}
{\bf Proof.}
The expansion (\ref{eq:expmar}) is just that of Theorem~\ref{thwg}. Note that 
by $\ker P(s)=\ran L(s)$ (or, more abstractly, by Lemma~\ref{lem:1dimS} for
$\mathcal{A}=\ell^\infty(S)$) 
the hypothesis of Corollary~\ref{thm:cor1} is satisfied. Thus $T(s,\,s') \pi(s') = \pi(s)$ and $a_1(s) = 0$. \hfill$\square$\\



We say that $L$ satisfies a {\em detailed balance} if   
\begin{equation}
M_{ij}:=L_{ij} \pi_j 
\label{eq:db}
\end{equation}
is a symmetric matrix for some $\pi$, in which case that is the stationary
distribution. This can be interpreted as the statement that the current through any link $j\to i$
\begin{equation}\label{eq:pcrt}
J_{ij}(p)= L_{ij} p_j-L_{ji}p_i
\end{equation}
vanishes at equilibrium, $J_{ij}(\pi)=0$.

We now strengthen the assumption on $S$ from indecomposable to
irreducible, meaning that $\pi_j>0$. Then $\ker M=\lsp\{(1,1,\ldots 1)\}$, 
$\ran M=\ran L$, and the two subspaces decompose $M$ as a linear map. 
At first $M^{-1}$ is defined 
on $\ran M$, and it may be extended afterwards, arbitrarily but linearly,
to all of $\mathbb{C}^d$, e.g. by having it vanish on $\ker M$.
 
In applications to (stochastic) molecular pumps one is interested in systems that carry no current in
their equilibrium states, but can be induced to yield net particle transport
in an adiabatic pump cycle.
Note first that $M$ and $\pi$ provide natural coordinates for 
those irreducible processes $L$ which satisfy a detailed balance condition.
We set the pump period (in scaled time) to be unity.  
 
The net transport across the link $j\to i$ is expressed in terms of the integrated probability current 
\begin{equation*}
 T_{ij} := \frac{1}{\varepsilon}\int_0^1 J_{ij}\big(p(s)\big)\, ds.
\end{equation*}
The following describes the current in the adiabatic limit.
	\begin{cor}
	Let $s\mapsto \{M(s),\pi(s)\}$ be a pump cycle with $\pi(s)$ the unique equilibrium state for every $s$. Assume that $\dot\pi_j(0) = 0$. Then the transport is geometric to leading order, given by
	\begin{align}
	T_{ij} 
	&= \int_0^1 \sum_{k=1}^d \left( M_{ij}(s) M^{-1}_{jk}(s)
	-M_{ji}(s)M^{-1}_{ik}(s) \right) d\pi_{k}(s)\, + O(\varepsilon).
	\label{eq:cur}
	\end{align}
	In particular $T_{ij} = O(\varepsilon)$ if $\pi$ is constant or, in the periodic
	case $L(0)=L(1)$, if $M$ is.
	\end{cor}
\begin{rem} Here, geometric means that the transport  is independent of the parametrization of the pumping cycle. This is evident in Eq.~(\ref{eq:cur}).
\end{rem}
\begin{rem}
The corollary says that  {\em effective pump cycles} require the variation of both  $\pi$ and $M$. As a matter of fact, the long time average of $T_{ij}$ vanishes under the conditions stated in the last line of the corollary regardless of the adiabatic limit \cite{RHJ08, Maes, JM11}.
\end{rem}

\noindent
{\bf Proof.} The contribution to $T_{ij}$ of order $\varepsilon^{-1}$
vanishes due to the detailed balance condition. 
To next order Eqs.~(\ref{eq:pcrt}, \ref{eq:expmar}) yield
\begin{equation*}
T_{ij} =\int_0^1 \sum_{k=1}^d \bigl( L_{ij}(s) L^{-1}_{jk}(s)
	-L_{ji}(s)L^{-1}_{ik}(s) \bigr) \dot\pi_{k}(s)\, ds + O(\varepsilon).
\end{equation*}
Eq.~(\ref{eq:db}) may be written as
$M=(1-P)L(1-P)\Pi$, where $\Pi_{ij}=\pi_i\delta_{ij}$, implying
$L^{-1}=(1-P)\Pi M^{-1}(1-P)$. Thus, with $P_{jl}=\pi_j$, we have
\begin{equation*}
L_{ij} L^{-1}_{jk}\dot\pi_{k}=
L_{ij}\sum_l (1-P)_{jl}\pi_l M^{-1}_{lk}\dot\pi_{k}=
M_{ij}M^{-1}_{jk}\dot\pi_{k} - \sum_l M_{ij}\pi_l M^{-1}_{lk}\dot\pi_{k}.
\end{equation*}
After interchanging $i, j$ and taking the difference, the second term cancels
and we are left with (\ref{eq:cur}). The additional
claim in the periodic case follows by the fundamental theorem of 
calculus.\hfill$\square$

\subsection{Remarks about the gapless case}\label{subsec:gapless}
Theorem~\ref{thwg2} can be applied to evolutions generated by either a
Hamiltonian or a Lindbladian, just like Theorem~\ref{thwg} was in 
Subsecs.~\ref{subsec:uev} and \ref{subsec:thermal}, respectively.
The Hilbert space $\mathcal{H}$ must, of course, be infinite-dimensional in order for the gap to vanish.  

In the unitary case the result provides a new proof of the adiabatic theorem for Hamiltonians without spectral gap, as in \cite{AE99, Te01}; in fact,
Hypotheses \ref{h1} and \ref{h2prime}' are trivially satisfied in this case.

In the Lindbladian case it would be desirable to apply Theorem~\ref{thwg2} to
the natural space $\mathcal{B}=\mathcal{J}_1(\mathcal{H})$. Unfortunately its Hypothesis~\ref{h2prime}' is typically not satisfied, as Example~\ref{exlind}
shows. However, Lindbladians of the dephasing kind have extensions to
$\mathcal{J}_2(\mathcal{H})$, to which that hypothesis does
apply. The result so
obtained is applicable to tunneling, as we shall see below.  
A typical situation leading to a gapless (dephasing) Lindbladian arises from a
Hamiltonian having both continuous and point (e.g. discrete) spectrum. 

\begin{thm} \label{thwgopen} 
Let $\mathcal{L}(s)$, $0 \leq s \leq 1$, be $C^1$-family of dephasing
Lindbladians acting on $\mathcal{J}_2(\mathcal{H})$. Then Eq.~(\ref{eq:tr2})
holds true. Let $\mathcal{P}(s)$ be the
associated projections, for almost all $s$; moreover let $\mathcal{P}(s)$ be defined for all $0 \leq s \leq 1$ and $C^1$ as a bounded operator on $\mathcal{J}_2(\mathcal{H})$. Then the solution of $\varepsilon \dot{\rho}=\mathcal{L}(s)\rho$ with initial data $\rho(0)=\mathcal{P}(0)\rho(0)$ satisfies
\begin{align}
\sup_{s\in [0,1]} \|\rho(s)-\mathcal{T}(s,0)\rho(0)\|_{\mathcal{J}_2(\mathcal{H})} \to 0,\qquad (\varepsilon \to 0).
\end{align}
\end{thm}
The assumptions on the dephasing Lindbladian $\mathcal{L}(s)$ follow from
corresponding ones on the underlying Hamiltonian $H(s)$, see
Eq.~(\ref{eq:dldef}): Let $P_j(s)$ be its eigenprojections, for almost all
$s$, cf.~Remark~\ref{rem:aa}, and let them be defined for all 
$0 \leq s \leq 1$ and $C^1$ uniformly in $j$. 
Then, as will be proved together with the theorem, 
$\mathcal{T}(s,0)P_j(0)=P_j(s)$. This implies the following result about
tunneling out of an initial state given by a rank 1 eigenprojection
$\rho(0)=P_0(0)$:
\begin{equation*}
T_{\varepsilon}(s)= 1-\tr(\rho(s)P_0(s))=
1-\tr((\mathcal{T}(s,0)P_0(0))P_0(s))+o(1)=o(1).
\end{equation*}

The next example illustrates that, although the tunneling out of 
the initial state is of order $o(1)$, an adiabatic invariant is conserved 
up to order $\varepsilon$.
\begin{exa}
Consider the Hamiltonians $H(s) = V(s) H V^*(s)$ arising from a $C^2$-family 
of unitaries $V(s)$ and from a bounded $H$. Let 
$\rho(s)$ solve the equation $\varepsilon \dot \rho = -\im[H,\rho]$. Then the
energy is an adiabatic invariant in the sense 
that 
$$
\bigl|\tr(H(1)\rho(1)) - \tr(H(0) \rho(0))\bigr| = O(\varepsilon).
$$
This follows from Eq.~(\ref{adinv2}). We may in fact apply that estimate to
$x(s)=\rho(s)$, $\varphi(s)=H(s)$, $\langle \varphi$, $x \rangle=\tr(H\rho)$ and
${\mathcal L} =-\im[H,\cdot]$, since the assumptions (\ref{phicond2}) hold true by 
\begin{equation*}
{\mathcal L}^*(s)(H(s))=0,\qquad\dot H(s) = -[H(s), \dot V(s) V^*(s)]=
{\mathcal L}^*(s)(\im\dot V(s) V^*(s)).
\end{equation*}
\end{exa}

\section{Proofs and supplementary results}\label{proofs}

We begin by recalling the Hille-Yosida theorem (\cite{RS2}, Thm. X.47a): 
A densely defined, closed operator 
$L$ on $\mathcal{B}$ generates a contraction semigroup iff 
\begin{equation}
\label{eq:hy}
\begin{gathered}
(0,\infty)\subset \rho(L),\\ 
\|(L-\gamma)x\|\ge \gamma\|x\|,\qquad
(\gamma>0, x\in D(L)).
\end{gathered}
\end{equation}
Conditions (\ref{eq:hy}) reflect the connection between the resolvent
and evolution operators. For example the only if part of Hille-Yosida theorem
 follows from the formula
$$
-(L - \gamma)^{-1} = \int_0^\infty \e^{(L - \gamma)t} d t, \qquad 
(\gamma > 0).
$$

\noindent
{\bf Proof of Lemma~\ref{lem:ev}.} The hypotheses are a convenient 
strengthening of those of \cite{RS2}, Thm.~X.70, including the remark 
thereafter. All our statements but uniqueness and 
Eq.~(\ref{eq:ev2}) are among its claims, and those two are consequences of 
its proof.
Alternatively, the results may be read off from \cite{K3}:
Eq.~(\ref{eq:ev1}) from Thm.~4, Eq.~(\ref{eq:ev1a}) and uniqueness from Thm.~1, and Eq.~(\ref{eq:ev2}) from Thm.~2.\hfill$\square$\\

\comment{
{\tt for internal use}
{\bf Proof of Lemma~\ref{lem:ev}.} The parameter $\varepsilon$ may be absorbed
in $L$ without loss. The hypotheses are a convenient strengthening of those
of (\cite{RS2}, Thm. X.70), including the remark thereafter. All our statements but (\ref{eq:ev2}) are among its
claims, and that one is a consequence of its proof, as we will show. Actually,
the proof in \cite{RS2} makes the additional assumption that $0\in \rho(L(s))$
which, as remarked there, can be arranged for by replacing $L(s)$ with 
$L(s)-c$, ($c>0$). It is then shown that
\begin{equation*}
(U(s'',s')-1)x=\int_{s'}^{s''}W(r,s')L(s')x\,dr\,,\qquad
(x\in D)\,,
\end{equation*}
where $W(r,s')=L(r)U(r,s')L(s')^{-1}$ is a bounded operator on $\mathcal{B}$,
jointly strongly continuous in $r$, $s'$. In particular, $W(r,r)=\id$. 
Let $0\le s\le 1$, $x\in D$
and $\varepsilon>0$ be given. For all $r, s'\in [s-\delta, s+\delta]\cap
[0,1]$ we have
\begin{equation*}
 W(r,s')L(s')x-L(s)x=W(r,s')(L(s')-L(s))x
+(W(r,s')-1)L(s)x\,,
\end{equation*}
which by (ii) is estimated in the norm as
$C\|(L(s')-L(s))x\|+\|(W(r,s')-1)L(s)x\|\le\varepsilon$, provided 
$\delta>0$ is small enough. Hence, for $s''\ge s\ge s'$,
\begin{equation*}
(U(s'',s')-1)x=(s''-s')L(s)x+o(s''-s')\,,\qquad(s''-s'\to 0)\,.
\end{equation*}
Eq.~(\ref{eq:ev2}) then follows by the group property of
$U$. \hfill$\square$\\}

\noindent
{\bf Proof of Lemma~\ref{lem:1dim}.}
Rank 1 projections $P$ are of the form $Py=\alpha(y)x$
where $x\in \mathcal{B}$ and $\alpha\in\mathcal{B}^*$ are determined up to
reciprocal factors. Any $\alpha$ with $\ker\alpha=\ker P$ may thus
be picked, and then $x$ normalized by $\alpha(x)=1$. Since $\ker P(s)$ is independent of $s$, so is our choice of
$\alpha$ in $P(s)y=\alpha(y)x(s)$, while 
$x(s)$ is $C^1$. Thus $\dot{P}(s)y=\alpha(y)\dot x(s)$, which vanishes for 
$y\in \ker P(s)$. The claim just proved states $\dot{P}=\dot{P}P$; together
with (\ref{eq:pt2}) both sides of $P(s)=T(s,s')P(s')$ are seen to satisfy the
same differential equation in $s$.\hfill$\square$\\

Consider the ranges of $P(s)$ and of $Q(s)=\id -P(s)$. As the name suggests, parallel transport $T(s,0)$ is obtained by projecting
vectors from either subspace at $0$ to the corresponding one at $s$ or, more
precisely, by repeating the procedure on the intervals of an ever finer
partition of $[0,s]$. In fact,
\begin{equation*}
P(s)P(0)+Q(s)Q(0)=\id+[\dot{P}(0),P(0)]s+o(s), \qquad (s\to 0),
\end{equation*}
implying by Eq.~(\ref{partra})
\begin{align}
T(s,s')&=\lim_{N\to\infty}
\prod_{i=0}^{N-1}(P(s_{i+1})P(s_i)+Q(s_{i+1})Q(s_i))\nonumber\\
&=\lim_{N\to\infty}\bigl(\prod_{i=0}^{N}P(s_i)+\prod_{i=0}^{N}Q(s_i)\bigr),
\label{eq:ptm1}
\end{align}
where $s'=s_0\le s_1\le\ldots\le s_N=s$ is a partition of $[s',s]$ into
intervals of length $|s_{i+1}-s_i|=N^{-1}|s-s'|$ and
$\prod_{i=0}^{N-1}A_i=A_{N-1}\cdots A_0$.\\

\noindent
{\bf Proof of Proposition~\ref{lem:NeumannTr}.} 
Since the product of state preserving maps is state preserving
the first claim follows from (\ref{eq:ptm1}). As a result 
$\mathcal{T}(s,s')\mathcal{P}(s')$ maps ${\mathcal S}(s')\to {\mathcal S}(s)$, with inverse
$\mathcal{T}(s',s)\mathcal{P}(s)$. The first bullet follows from Eq.~(\ref{eq:kad}) for the two 
maps, i.e.
$$ \|\rho(s')-\tilde\rho(s')\|=\|\mathcal{T}(s',s)\mathcal{P}(s)(\rho(s)-\tilde\rho(s))\|\le \|\rho(s)-\tilde\rho(s)\|.$$
Any convex 
decomposition of $\mathcal{T}(s,s')\rho(s')$, ($\rho(s')\in {\mathcal S}(s')$) entails one 
of $\rho(s')$, which yields the second bullet in the variant where the bracketed
word is omitted. The continuity of $\mathcal{T}(s,s')\rho$ w.r.t. $\rho$ yields the other
variant. To obtain the last statement we note that 
$\mathcal{T}(s,s')\rho(s')$ is, for fixed $s$ and all $s'$, an 
isolated extreme point in ${\mathcal S}(s)$, just like $\rho(s)$. 
They agree, since they do for $s'=s$.
\hfill$\square$\\

\noindent
{\bf Proof of Lemma~\ref{lem:1dimS}.} 
By the setting of Subsec.~\ref{subsec:states}, $\mathcal{P}$ has a predual 
$\mathcal{P}_*$, which is also of rank $1$. Since
$\mathcal{P}_*(\id)=\id$ by the normalization condition, we have 
$\ran \mathcal{P}_*=\lsp\{\id\}$. Thus 
$\ker\mathcal{P}=(\ran \mathcal{P}_*)^\perp$ is independent of $\mathcal{P}$. 
We recall that $S^\perp\subset \mathcal{B}=\mathcal{A}^*$ is the annihilator 
of a subspace $S\subset \mathcal{A}$. The remaining claims follow from
Lemma~\ref{lem:1dim} and Proposition~\ref{lem:NeumannTr}.\hfill$\square$\\ 

\noindent
{\bf Proof of Proposition~\ref{prop:directSum}.} $La=0$ and $a=Lb$ imply $(L-\gamma)(a+\gamma b)=-\gamma^2 b$ and 
by (\ref{eq:hy}) $\gamma\|a+\gamma b\|\le \gamma^2 \|b\|$ for $\gamma>0$. 
After dividing by $\gamma$ we obtain $a=0$ in the limit $\gamma\to 0$. \hfill $\square$\\

Before proving Theorem~\ref{thwg} we derive a few further consequences of its
assumptions.
Hypothesis~\ref{h2} includes Eq.~(\ref{eq:inter0}); moreover the pair of subspaces
decomposes $L$. The
restriction $L\restriction\ran L$ is closed and has range 
$\ran(L\restriction\ran L)=\ran L$; by (\ref{eq:inter0}) it is one-to-one. 
Thus 
$0\notin\sigma(L\restriction\ran L)$. Together with 
$\sigma(L\restriction\ker L)\subset\{0\}$ we conclude that the resolvent set 
contains a punctured neighborhood of $0$, which proves the presence of a gap.

We maintain that the projection $P$ is given by the Riesz projection, see 
Eq.~(\ref{eq:rp}) with $\lambda=0$. Calling the latter temporarily $\tilde P$
we thus claim $\tilde Pa=a$ for $a\in \ker L$ 
and $\tilde Pb=0$ for 
$b\in\ran L$. The first statement is evident from (\ref{eq:rp}); for
the second it suffices, by $\tilde PL\subset L\tilde P$, to show that
$\ran \tilde P\cap \ran L=\{0\}$. This in turn follows because 
$L\restriction(\ran \tilde P\cap \ran L)$ is a bounded operator with empty 
spectrum; in fact, it is contained in 
$\sigma(L\restriction\ran \tilde P)\cap\sigma(L\restriction\ran L)=\varnothing$
since the first spectrum is contained in $\{0\}$, while the second is disjoint
from it. Finally, we recall the formula for the inverse of
$L\restriction\ran(1- P)$ (\cite{K2}, Eq.~(III.6.23)):
\begin{equation*}
L^{-1}=-\frac{1}{2\pi\im}\oint_\Gamma(L-z)^{-1}\frac{dz}{z}.
\end{equation*}
In particular, $P(s)$ and $L(s)^{-1}$ are $C^k$ in norm. \\

\noindent
{\bf Proof of Theorem~\ref{thwg}.}
We insert the right-hand side of (\ref{solu}) as an ansatz into (\ref{adeq})
and equate orders $\varepsilon^n, \ (n=0,\dots, N)$, resp. 
$O(\varepsilon^{N+1})$. We find
\begin{align}
Lb_0&=0, \nonumber \\
\dot{a}_n+\dot{b}_n&=Lb_{n+1}, \qquad (n=0, \dots, N-1) \label{rec_1} \\
\varepsilon \dot{r}_N+\dot{a}_N+\dot{b}_N&=Lr_N. \label{rec_2}
\end{align}
In particular, $b_0=0$. Note that $Qa=0$ implies $\dot{Q}a+Q\dot{a}=0$, or
$Q\dot{a}=-\dot{Q}a=\dot{P}a$. Similarly, $P\dot{b} = -\dot{P}b$. Applying $Q$
and $P$ to (\ref{rec_1}) yields
\begin{align}
 \dot{P}a_n+Q\dot{b}_n&=Lb_{n+1} \label{intermed_1}, \\
P\dot{a}_n-\dot{P}b_n&=0. \label{intermed_2}
\end{align}
If $b_n$ is known, (\ref{intermed_2}) implies
\begin{align*}
 \dot{a}_n=Q\dot{a}_n+P\dot{a}_n=\dot{P}a_n+\dot{P}b_n,
\end{align*} 
the solution of which is (\ref{an}) by Eq.~(\ref{eq:pt2}) and the Duhamel
formula. If $a_n$ and $b_n$ are known, $b_{n+1}$ follows from
(\ref{intermed_1}) and Lemma~\ref{lem:tr}, provided $b_n$ is differentiable (see below). All 
this determines 
$b_0, a_0, b_1,\dots, a_{N-1},b_N$. We
then define $a_N, b_{N+1}$ by the same Eqs. (\ref{an}, \ref{bnpo}), which ensures $\dot{a}_N+\dot{b}_N=Lb_{N+1}$. Then (\ref{rec_2}) reads
\begin{align*}
 \varepsilon \dot{r}_N =Lr_N-Lb_{N+1}
\end{align*}
with solution
\begin{align*}
 r_N(\varepsilon,s)&=U_\varepsilon(s,0)r_N(\varepsilon,0)-\varepsilon^{-1}\int_0^sU_\varepsilon(s,s')L(s')b_{N+1}(s')ds' \\
&=U_\varepsilon(s,0)r_N(\varepsilon,0)+\int_0^s(\frac{\partial}{\partial s'}U_\varepsilon(s,s'))b_{N+1}(s')ds'.
\end{align*}
An integration by parts yields (\ref{rN}) and the bound on the remainder
follows from Hypothesis \ref{h1} through Eq.~(\ref{eq:ev1a}). Inspection of the recursion
relations shows $a_n,b_n\in C^{N+2-n}$, which provides the required
differentiability. 
\hfill$\square$\\

\noindent
{\bf Proof of Theorem~\ref{thm:cor2}.} For an adiabatic invariant $\varphi$ 
Eq.~(\ref{adinv1}) together with 
	$\langle \varphi(0), x(0)\rangle=\langle \varphi(s), T(s,0)x(0)\rangle$ yields
	\begin{equation*}
	|\langle\varphi(s), x(s)-T(s,0)x(0)\rangle|\le 
	C\varepsilon\|\varphi(s)\|\|x(0)\|.
	\end{equation*}
	The claim follows from 
	$\|P(s)x\|=\sup\{|\langle\varphi,x \rangle|\mid
	\varphi\in \ker L^*(s), \|\varphi \|=1\}$. \hfill$\square$\\

\noindent
{\bf Proof of Proposition~\ref{lm:adinv}.} The assumption (\ref{phicond}) may
	be phrased differently. The projections $P^*$ and $Q^*$ are associated to 
	$\ker L^* \oplus \ran L^*$, because the subspaces (\ref{eq:tr}) decompose $L$. Thus $Q^*\varphi=0$, $P^*\dot{\varphi}=0$ just 
	means that $\varphi(s)\in\ran P^*(s)$ is parallel transported: 
	$\varphi(s)=T^*(s,s')\varphi(s')$.

	Eq. (\ref{adinv}) follows from  
 	\begin{align*}
 	\frac{d}{ds}\langle \varphi, x \rangle &= \langle \dot{\varphi},x \rangle+ \langle \varphi, \dot{x} \rangle 
	=\langle L^*{L^*}^{-1}\dot{\varphi},x \rangle + \varepsilon^{-1}\langle \varphi, L x \rangle \\
	&= \langle {L^*}^{-1}\dot{\varphi}, Lx \rangle + \varepsilon^{-1} \langle L^* \varphi, x \rangle 
	= \varepsilon \langle {L^*}^{-1} \dot{\varphi}, \dot{x} \rangle + 0\,.
 	\end{align*}
	Integration by parts in (\ref{adinv}) gives 
	\begin{equation}\label{eq:iparts}
 	\langle \varphi, x \rangle |_0^s = \varepsilon 
	\Bigl( \langle\phi,x\rangle|_0^s- 
	\int_0^s\langle\dot{\phi}(s'),x(s')\rangle ds'\Bigr),
	\end{equation}
	where $\phi(s')={L^*(s')}^{-1}\dot{\varphi}(s')$.
	We observe that $\|x(s')\|\le \|x(0)\|$ by Lemma~\ref{lem:ev}, and 
	$\dot{\varphi}(s)=\dot{P}^*(s)\varphi(s)$ by (\ref{eq:pt2}). By 
	(\ref{eq:trbound}) we see that $\|\varphi(s')\|$, $\|\phi(s')\|$ and 		$\|\dot{\phi}(s')\|$ are bounded by a constant times 
	$\|\varphi(s)\|$, proving Eq.~(\ref{adinv1}).
	\hfill$\square$\\

\noindent
We now turn to the case without gap. \\

\noindent
{\bf Proof of Theorem~\ref{thwg2}.}
Let us first dispose of the ``almost all" qualifier in the theorem.
We observe that, by continuity, $L(s)P(s)=0$ holds for all $0\le s\le 1$. In 
particular, $L(s)x_0(s)=0$ for $x_0(s)=T(s,0)x(0)$. The remainder to be
estimated is $r(s)=x(s)-x_0(s)$. By Eq.~(\ref{adeq}) it satisfies the 
differential equation 
$\varepsilon \dot r(s)=L(s)r(s)-\varepsilon\dot{x}_0(s)$ with solution 
\begin{equation*} 
r(s)=-\int_0^s {U}_{\varepsilon}(s,s')\dot{x}_0(s')ds'.
\end{equation*} 
By Eqs.~(\ref{eq:pt}, \ref{eq:tr2}) we have 
$\dot{x}_0(s)\in \overline{\ran L(s)}$ for almost all $s$. This property is
reflected in the splitting 
\begin{equation*} 
\dot{x}_0(s)=L(s)(L(s)-\gamma)^{-1}\dot{x}_0(s)-
\gamma(L(s)-\gamma)^{-1}\dot{x}_0(s)\,,\qquad (\gamma>0)\,,
\end{equation*} 
motivated by Eq.~(\ref{eq:eqr}) below. Together they yield
\begin{equation*} 
r(s)=-\int_0^sU_{\varepsilon}(s,s')L(s')(L(s')-\gamma)^{-1}\dot{x}_0(s')ds'+
\int^s_0U_{\varepsilon}(s,s')\gamma(L(s')-\gamma)^{-1}\dot{x}_0(s')ds',
\end{equation*} 
where, by an appropriate choice of $\gamma>0$, the second integral can be made 
arbitrarily small by means of dominated convergence; in fact, uniformly in
$\varepsilon$ due to $\|U_{\varepsilon}(s,s')\|\le 1$. It remains to show
that, for fixed $\gamma>0$, the first integral vanishes with $\varepsilon$.
To illustrate the argument, let us temporarily pretend that 
$z(s):= (L(s)-\gamma)^{-1}\dot{x}_0(s)$ is $C^1$. Since 
$\varepsilon \partial_{s'} U_\varepsilon(s,s')= -U_\varepsilon(s,s')L(s')$
an integration by parts yields for that integral
\begin{equation*} 
\varepsilon\int_0^s\partial_{s'}{U}_{\varepsilon}(s,s')z(s')ds'=\varepsilon{U}_{\varepsilon}(s,s')z(s')|_{s'=0}^{s'=s}-\varepsilon \int_0^s{U}_{\varepsilon}(s,s')\frac{d}{ds'}z(s')ds',
\end{equation*} 
and exhibits the desired property for $\varepsilon\to 0$. Finally, we get 
rid of the additional assumption by amending the argument as follows. 
We introduce a
mollifier $j$, ($j\in C_0^\infty(\mathbb{R})$, $\int j(x)dx=1$) and set 
$j_\delta(x)=\delta^{-1}j(x/\delta)$, ($\delta>0$); we extend $\dot{x}_0$ 
continuously outside of the interval $[0,1]$; and split
\begin{equation*} 
z=(L-\gamma)^{-1}(\dot{x}_0-j_\delta*\dot{x}_0)
+(L-\gamma)^{-1}(j_\delta*\dot{x}_0). 
\end{equation*} 
Since $\dot{x}_0-j_\delta*\dot{x}_0\to 0$, ($\delta\to 0$) 
and $\|L(L-\gamma)^{-1}\|$ is bounded, both uniformly in $s$, the first term
contributes arbitrary little to the integral, uniformly in $\varepsilon$, if
$\delta$ is picked small enough. The preliminary argument can now be 
applied to the second term in place of $z$. \hfill$\square$\\

\noindent
{\bf Proof of Proposition~\ref{lm:adinv1}.} Eq.~(\ref{eq:iparts}) can be
obtained from the present assumptions by replacing $\phi$ for 
${L^*}^{-1}\dot{\varphi}$ in the previous
derivation.\hfill$\square$\\

\noindent
{\bf Proof of Lemma~\ref{lem:tr}.} Suppose a closed operator $L$ has $0$
as an isolated point in its spectrum, with associated Riesz projection 
$P$, see Eq.~(\ref{eq:rp}) with $\lambda=0$. Then $P$ decomposes $L$ with $\sigma(L\restriction\ran
P)=\{0\}$
and $\sigma(L\restriction\ran (1-P))=\sigma(L)\setminus\{0\}$ 
(\cite{K2}, Thm. III.6.17), whence $\ran L\supset\ran(1-P)$ and 
$\ker L\subset\ran P$. As a result, (H2i) implies that $L$ is Fredholm because
both its parts are, and in particular that (H2ii) holds, regardless of Hypothesis \ref{h1}. 

From now on we assume (H1) and (H2ii), with the former implying
Eqs.~(\ref{eq:hy}, \ref{eq:inter0}). 
By the stability theorem (\cite{K2}, Thm. IV.5.31) 
$L-z$ remains semi-Fredholm for $z$ in a
complex neighborhood of $0$ and the index 
$\dim\ker(L-z)-\dim(\mathcal{B}/\ran(L-z))$ is constant; moreover the two 
dimensions are
separately constant in a punctured neighborhood $U$. By (\ref{eq:hy}), they both
vanish there, and so does the index at $z=0$. This has the following
implications: First, if $0\in\sigma(L)$, then it is isolated. Second, the
map $\ker L\to\mathcal{B}/\ran L$, $a\mapsto a+\ran L$ is one-to-one
by (\ref{eq:inter0}) and thus onto by the vanishing index. This proves
$\ker L+ \ran L=\mathcal{B}$, completing the proof of Eq.~(\ref{eq:tr}). 

Finally in order to prove (H2i), we observe that the Riesz projection 
is given by $P$, as established at the beginning of this section. Thus it 
is finite-dimensional because $\ker L$ is.\hfill$\square$\\

\noindent
Lemma~\ref{lem:tr2} is an immediate consequence of the last two statements of
the following lemma.
\begin{lem} 
\label{lem:tr3}
Let $L$ be the generator of a contraction semigroup on $\mathcal{B}$. Then 
\begin{gather}
\label{eq:eqr}
b\in \overline{\ran L}\; \Leftrightarrow\;
 \lim_{\gamma \downarrow 0} \gamma(L-\gamma)^{-1}b=0,\qquad 
(b\in\mathcal{B}),\\
\label{eq:eqr2}
\ker L\cap \overline{\ran L}=\{0\}\
\end{gather}
If, in addition, $\mathcal{B}$ is reflexive, then 
$\overline{\ker L+\ran L}=\mathcal{B}$.
\end{lem}

\noindent
{\bf Proof.} We note that $\gamma(L-\gamma)^{-1}$ is uniformly bounded in 
$\gamma>0$ by (\ref{eq:hy}). 
By density, it
thus suffices to prove the direct implication (\ref{eq:eqr}) for $b=Lx$, 
for which it follows
from $\gamma(L-\gamma)^{-1}Lx=\gamma(x+\gamma(L-\gamma)^{-1}x)$. Conversely,
set $x_\gamma=(L-\gamma)^{-1}b$; then 
$Lx_\gamma=b+\gamma(L-\gamma)^{-1}b\to b$.

Next, let $b$ be in the intersection (\ref{eq:eqr2}): we have 
$(L-\gamma)b=-\gamma b$,
and hence $b=-\gamma(L-\gamma)^{-1}b$, which vanishes as $\gamma \downarrow 0$.

To prove the last claim it suffices, by the Hahn-Banach theorem, to show
that $x^*\in(\ker L)^\perp\cap(\ran L)^\perp$ implies $x^*=0$. Here
$S^\perp\subset \mathcal{B}^*$ is the annihilator of a subspace 
$S\subset \mathcal{B}$. We have $(\ran L)^\perp=\ker L^*$ and, in the
reflexive case,
$(\ker L)^\perp=\overline{\ran L^*}$. The last equality is due to 
$S^{\perp\perp}=\overline{S}$ (\cite{K2}, Eq. III.1.24). The property 
$(0,\infty)\subset \rho(L)$ and the uniform bound on 
$\gamma(L-\gamma)^{-1}$, ($\gamma>0$) are inherited by $L^*$, and so is the
consequence (\ref{eq:eqr2}). We conclude that $x^*=0$.
(As a matter of fact, $L^*$ is also densely defined (\cite{K2} Thm. III.5.29)
by reflexivity, and hence is the generator of a contraction semigroup; in
particular $(\e^{Lt})^*=\e^{L^*t}$.)
\hfill$\square$\\

This concludes the proofs of Sect.~\ref{sec:res} and we pass to those of
Sect.~\ref{sec:appl}. Actually, Theorems~\ref{expTunneling} and \ref{LindTunneling}
do not require proof, as they are immediate applications of Sect.~\ref{sec:res}, while the results of Subsec.~\ref{subsec:nick} are proven there.
\\

\noindent
{\bf Proof of Proposition~\ref{lm:deph}.} To begin we write Eq.~(\ref{eq:lin}) as 
\begin{equation*}
\mathcal{L}\rho=-\im[H,\rho]+\frac 1 2 \sum_\alpha\bigl( 2\Gamma_\alpha \rho
\Gamma_\alpha^*-\{\rho,\Gamma_\alpha^*\Gamma_\alpha\}\bigr),
\end{equation*}
and hence 
\begin{equation*}
\mathcal{L}_*a=\im[H,a]+\frac 1 2 \sum_\alpha \bigl(2\Gamma_\alpha^* a
\Gamma_\alpha-\{a,\Gamma_\alpha^*\Gamma_\alpha\}\bigr).
\end{equation*}
It is evident that $\Gamma_\alpha=f_\alpha(H)$ implies 
$\ker \mathcal{L}_*\supset\ker([H,\,\cdot])$ and, through 
$[\Gamma_\alpha^*,\Gamma_\alpha]=0$, also 
$\ker \mathcal{L}\supset\ker([H,\,\cdot])$.
Together with a detail to be supplied momentarily, the three claims thus reduce to the following ones:
\begin{enumerate} 
\item 
$\ker \mathcal{L}_*\supset
\ker([H,\,\cdot])$ implies 
$\Gamma_\alpha\in \{f(H)\}''\equiv \{f(H)\mid f(H)\in\mathcal{B}(\mathcal{H})\}''$. 
\item 
$\Gamma_\alpha=f_\alpha(H)$ implies 
$\ker \mathcal{L}\subset
\ker([H,\,\cdot])$.
\item 
If the spectrum of $H$ is pure point and if $\ker
\mathcal{L}\supset
\ker([H,\,\cdot])$, then $\Gamma_\alpha\in \{f(H)\}''$. 
\end{enumerate} 
Here a prime denotes the commutant. We recall (\cite{BR}, Thm. 2.4.11) that
the bicommutant is the strong closure of the original $*$-algebra. Note that 
$\{f(H)\}$ is strongly closed.

The implications 1.-3. are based on the readily verified identity \cite{Li76}
\begin{equation}
\label{diss}
\mathcal{L}_*(a^*a)-a^*\mathcal{L}_*(a)-\mathcal{L}_*(a^*)a=
\sum_\alpha[a,\Gamma_\alpha]^*[a,\Gamma_\alpha].
\end{equation}

1. Let $a\in\ker([H,\,\cdot])=\{f(H)\}'$. 
 Since that subspace of 
$\mathcal{B}(\mathcal{H})$ is closed under taking adjoints and products, each term in the
l.h.s. of (\ref{diss}) vanishes by the assumption of the present item, implying 
$\Gamma_\alpha\in \{f(H)\}''$.

2. Under the assumption, $\mathcal{L}$ acts as $\mathcal{L}_*$ under the
replacement $H\to-H$, $\Gamma_\alpha\to\Gamma_\alpha^*$. Since 
$\tr\mathcal{L}(\rho)=0$ for $\rho\in\mathcal{J}_1(\mathcal{H})$, 
Eq.~(\ref{diss}) implies
\begin{equation*}
-\tr\rho^*\mathcal{L}(\rho)-\tr\rho\mathcal{L}(\rho^*)=
\sum_\alpha\tr[\rho,\Gamma_\alpha^*]^*[\rho,\Gamma_\alpha^*].
\end{equation*}
Thus $\rho\in\ker \mathcal{L}$ implies $[\rho,\Gamma_\alpha^*]=0$ and, by
$\mathcal{L}(\rho^*)=\mathcal{L}(\rho)^*$, also $[\rho,\Gamma_\alpha]=0$. We
conclude $[H,\rho]=0$.

3. By the first assumption we can pick finite-rank projections $P_n$,
which are sums of eigenprojections of $H$ or of subprojections thereof, such
that $P_n\displaystyle\mathop{\to}^s 1$. In particular $[H,P_n]=0$.

If $a\in\mathcal{J}_1(\mathcal{H})$ then
$\mathcal{L}_*(a)\in\mathcal{J}_1(\mathcal{H})$ and, we claim, 
$\tr\mathcal{L}_*(a)=0$ by our second assumption. Indeed, it implies 
$\mathcal{L}(P_n)=0$ and hence
\begin{equation*}
\tr\mathcal{L}_*(a)=\lim_n\tr(\mathcal{L}_*(a)P_n)=\lim_n\tr(a\mathcal{L}(P_n))=0\,.
\end{equation*}
Let now $a\in\mathcal{J}_1(\mathcal{H})\cap \{f(H)\}'$. Then 
$\mathcal{L}(a)=0$ and 
$\tr(\mathcal{L}_*(a^*)a)=\tr(a^*\mathcal{L}(a))=0$. By taking the trace of 
Eq.~(\ref{diss}) we conclude $[a,\Gamma_\alpha]=0$. The conclusion extends to 
$a\in \{f(H)\}'$ since $P_na\displaystyle\mathop{\to}^s a$. This proves the claim.
\hfill $\square$\\ 

\noindent
{\bf Proof of Theorem~\ref{thm:lind}.}
Clearly,
\begin{align}
 \dot{\mathcal{P}}\rho&=\sum_i(\dot{P}_i\rho P_i+ P_i \rho \dot{P}_i), \nonumber \\
\mathcal{L}^{-1}E_{ij}&=\lambda_{ij}^{-1}E_{ij}\,, \qquad (i\neq j). \label{helprel}
\end{align}
We note that 
\begin{equation}\label{eq:pt3}
 \mathcal{T}(s,s')P_k(s')=P_k(s).
\end{equation}
In fact, the l.h.s. satisfies the differential equation (\ref{partra}), viz.
\begin{align*}
 \frac{d}{ds}\rho(s)&=\dot{\mathcal{P}}(s)\rho(s), \\
\rho(s') &=P_k(s'),
\end{align*}
and so does the r.h.s., since
\begin{align*}
 \dot{\mathcal{P}}P_k=\sum_i\dot{P}_iP_kP_i+P_iP_k\dot{P}_i
=\dot{P}_k P_k +P_k\dot{P}_k=\dot{P}_k\,.
\end{align*}
The claim now follows from (\ref{exp1}-\ref{exp3}) with $a_0(0)=P_0(0)$. Indeed, the middle term of (\ref{Lindsol}) follows from (\ref{exp2}) and (\ref{helprel}):
\begin{align}
 \mathcal{L}^{-1}\dot{\mathcal{P}}P_0&= \mathcal{L}^{-1}\dot{P}_0= \mathcal{L}^{-1}\sum_{j\neq 0} P_j \dot{P}_0 P_0 +P_0 \dot{P}_0 P_j \nonumber 
\\&= \sum_{j\neq 0} \lambda_{j0}^{-1}P_j \dot{P}_0 P_0 +\lambda_{0j}^{-1}P_0 \dot{P}_0 P_j = 
\sum_{j\neq 0} \lambda_{j0}^{-1}P_j \dot{P}_0 +\lambda_{0j}^{-1} \dot{P}_0 P_j. \label{middle}
\end{align}
For the last term of (\ref{Lindsol}) we compute with (\ref{middle})
\begin{align*}
 \dot{\mathcal{P}}\mathcal{L}^{-1}\dot{\mathcal{P}}P_0&=\sum_{j\neq 0}\dot{\mathcal{P}}(\lambda_{j0}^{-1}P_j \dot{P}_0 P_0 +\lambda_{0j}^{-1}P_0 \dot{P}_0 P_j) \\
&=\sum_{j\neq 0}(\lambda_{j0}^{-1}+\lambda_{0j}^{-1})(P_0\dot{P}_j^2P_0 - P_j\dot{P}_0^2P_j ) \\
&=\sum_{j\neq 0}\alpha_j (P_j-P_0),
\end{align*}
(with termwise equality) where we have used $\dot{P}_iP_k = -P_i\dot{P}_k$ and
$\tr(P_j \dot{P}_0^2 P_j)=\tr(P_0\dot{P}_j^2P_0)$. Together with (\ref{exp3})
the expansion follows. The generalization follows because of the
contraction property of the propagator, Eq.~(\ref{eq:ev1a}).
\hfill$\square$\\

\noindent
{\bf Proof of Corollary~\ref{cor:lind}.} The statement evidently applies to
the ``slow manifold'' solution (\ref{Lindsol}), which however does not satisfy $\rho(0)=P_0(0)$, as required for the present solution. We compare the two by means of Remark~\ref{rem:cor2}. For their difference 
$\tilde\rho(s)$ we have $\|\tilde\rho(0)\|=O(\varepsilon)$ implying
$\|\mathcal{P}(s)\tilde\rho(s)\|=O(\varepsilon^2)$. 
\hfill$\square$\\

\noindent
{\bf Proof of Theorem~\ref{thwgopen}.} As noted at the beginning of
Subsec.~\ref{subsec:thermal} a Lindbladian $\mathcal{L}$ is the generator of a
contraction on $\mathcal{J}_1(\mathcal{H})$, while $\mathcal{L}_*$ is on 
$\com(\mathcal{H})$; for a dephasing Lindbladian the latter also applies to 
$\mathcal{L}$ itself, as noted in the proof of
Proposition~\ref{lm:deph}. Then, by interpolation, $\mathcal{L}$ 
generates a contraction also on $\mathcal{J}_2(\mathcal{H})$.

Given that $\mathcal{J}_2(\mathcal{H})$
is reflexive, Eq.~(\ref{eq:tr2}) follows from 
Lemma~\ref{lem:tr2} by showing that $\ker L+ \overline{\ran L}$ is
closed. We will actually show that the two subspaces are orthogonal w.r.t.
to the inner product $\langle\cdot,\cdot\rangle$ of 
$\mathcal{J}_2(\mathcal{H})$. In fact, for a dephasing Lindbladian the
statement $\ker \mathcal{L}=\ker([H,\,\cdot])$ from Prop.~\ref{lm:deph} also
holds true as subspaces of $\mathcal{J}_2(\mathcal{H})$, as inspection of the
proof shows. Thus $\ker \mathcal{L}\subset  \ker \mathcal{L}_*$ by 
(\ref{eq:dldef}). Then $\mathcal{L}a=0$ and $b=\mathcal{L}\tilde b$ imply
$\langle a,b\rangle=\langle \mathcal{L}_*a,\tilde b\rangle=0$. The theorem
follows.

To prove the statements following the theorem let us note that, for almost all $s$,
\begin{equation}\label{eq:kerpr}
\mathcal{P}(s)\rho=\sum_jP_j(s)\rho P_j(s),
\end{equation}
where the sum runs over the eigenvalues of $H(s)$. In fact, this
follows by the RAGE theorem (\cite{CFKS}, Thm. 5.8), i.e.
\begin{equation*}
\frac{1}{T}\int_0^T\e^{\im Ht}\rho\e^{-\im Ht}dt\to \sum_jP_j\rho
P_j,\qquad(T\to\infty)
\end{equation*}
for $\rho\in\com(\mathcal{H})$. It implies $\rho=\sum_jP_j\rho
P_j$ for $\rho\in\ker\mathcal{L}$. The converse is obvious. 
At this point the regularity of $\mathcal{P}(s)$ is inherited from that
of the $P_j(s)$, and Eq.~(\ref{eq:kerpr}) extends to all $s$.
Finally, $\mathcal{T}(s,0)P_j(0)=P_j(s)$ follows like
(\ref{eq:pt3}). \hfill$\square$\\

\medskip\noindent
{\bf Acknowledgements.} 
Example~\ref{exgang} is due to Gang Zhou. The results of Subsec.~\ref{subsec:nick} were derived in collaboration with Nicholas Crawford. We thank both of them for their contribution and for useful discussions.
JEA is supported by the ISF and the fund for promotion of research at the
Technion. MF is supported by the UNESCO. The authors benefitted from mutual visits supported by their
coauthors' institutions.

\bibliography{draft3.bib}
\bibliographystyle{plain.bst}


\end{document}